\documentclass[preprint,aps]{revtex4-1}
\usepackage{graphicx}
\usepackage{amssymb}
\usepackage{slashed}
\newcommand{\be}{\begin{eqnarray}}
\newcommand{\ee}{\end{eqnarray}}
\usepackage[colorlinks]{hyperref}
\usepackage{cancel}

\usepackage{amsmath,amssymb}
\usepackage{mathtools}
\usepackage{float}
\usepackage{color}
\usepackage{leftidx}
\usepackage{booktabs}
\usepackage{array}
\hypersetup{
    colorlinks=red,
    citecolor=blue,
    filecolor=magenta,      
    urlcolor=cyan,
}
\begin{document}

\title{Linearly polarized gluons in charmonium and bottomonium production in color octet model}

\author{Asmita Mukherjee and Sangem Rajesh}


\affiliation{ Department of Physics,
Indian Institute of Technology Bombay, Mumbai-400076,
India.}
\date{\today}

\begin{abstract}
We study the possibility to probe the 
unpolarized and linearly polarized transverse momentum-dependent gluon distributions 
 in unpolarized $pp$ collision in charmonium and bottomonium production,  employing non-relativistic QCD 
(NRQCD) based color octet model within transverse momentum dependent (TMD)
 factorization framework. The transverse momentum ($p_T$) and rapidity 
distributions  of $J/\psi$ and $\Upsilon(1\text{S})$ at LHCb, RHIC and AFTER energies are estimated. Significant modulations 
in transverse momentum spectrum of quarkonium in the low  $p_T$ region is obtained  when contributions of linearly
polarized gluons inside an unpolarized proton are taken into account. The results of quarkonium production
in color octet model and color evaporation model are compared.
\end{abstract}

\maketitle

\section{Introduction}\label{sec1}
Transverse momentum dependent (TMD) parton distribution functions (PDFs) and fragmentation functions (FFs), in short
TMDs  play an essential role in understanding the spin and spatial structure of the proton.  TMDs depend on both 
longitudinal momentum fraction $x$ and 
internal transverse momentum $k_\perp$ carried by partons in contrast to traditional collinear parton distributions (PDFs).
TMDs being nonperturbative objects have to be extracted from experimental data.
 The data mainly comes from two types of experiments, semi-inclusive deep inelastic scattering (SIDIS) and 
Drell-Yan (DY) process. In these experiments
one measures the transverse momentum of final observed hadron which is sensitive to the intrinsic transverse 
momentum ($k_\perp$) of the parton.\par
TMD factorization (generalized factorization) allows us to relate the cross section in terms of convolution of TMDs and 
hard scattering factor, which can be calculated in order of $\alpha_{s}$. TMD factorization has been studied extensively for 
SIDIS, DY and $e^+e^-$ annihilation process \cite{jcollins} at small transverse momentum ($p_T$) of the observed hadron
$i.e.$, $\Lambda_{QCD}\ll p_T\ll Q$, where $Q$ is the hard scale of the process. There are interesting issues 
related to the process dependence of the TMDs and the applicability of factorization for more general processes.
 Also an alternative approach to understand the single spin asymmetries  is based on collinear twist-3
  PDFs \cite{Kanazawa:2014nea,Gamberg:2014eia}.  At small transverse momentum, the radiative gluon 
emissions become important which need to be resummed up to all orders in $\alpha_{s}$. This is accomplished through 
TMD evolution
equation. Thus the TMDs satisfy a different and more involved evolution equation compared to the standard collinear PDFs.
 The evolution kernel is  calculated using perturbation theory which is independent of the process chosen and type of TMDs. 
 TMD evolution
equation, which can be obtained by solving renormalization group (RG) and Collins-Soper  \cite{jcollins} 
equations, is the consequence of TMD factorization. There is a non-perturbative part of the evolution kernel which is
 usually parametrized. \par
Within generalized factorization framework, the low $p_T$ region ($p_T\approx \mathcal{O}(10~\mathrm{GeV})$)
in differential cross section of $Z$-boson in $pp$ collision at $\sqrt{s}=8$ TeV (LHC) 
\cite{Angeles-Martinez:2015sea} and $\sqrt{s}=1.8$ TeV (CDF) \cite{Kulesza:2002rh} could be predicted to a good
accuracy. The reason is the resummation of large logarithmic corrections. 
However, the usual collinear factorization applied with convoluted collinear PDFs fails to describe
the low $p_T$ region of $Z$-boson spectrum. In this formalism,
radiative corrections make the cross section diverge as $p_T$ decreases \cite{Angeles-Martinez:2015sea}.
 In recent past, much work has been done to extract the quark TMDs from  experiments at COMPASS, HERMES 
and JLab \cite{Barone:2010zz,D'Alesio:2007jt}. However, very limited information is available about gluon TMDs
experimentally. To understand TMDs fully, we need global analysis of SIDIS and DY data. Nevertheless, the 
difficulty arises since the SIDIS and DY data span over different ranges in $p_T$ and  $\sqrt{s}$
\cite{Melis:2014pna}. In Higgs boson  production at NNLO \cite{denDunnen:2012ym}, the polarized gluons
contribute dominantly over the unpolarized gluons 
because of the fact that polarized gluons are generated from unpolarized gluons.\par
In Ref. \cite{Mulders:2000sh}, the authors pointed out that linearly polarized gluons can exist even at tree level inside an
unpolarized hadron. The associated density function 
denoted by $h_1^{\perp g}$, represents the probability of finding the linearly polarized gluon inside 
an unpolarized hadron.
For the existence of $h_1^{\perp g}$, gluons should have non-zero transverse momentum with respect to
hadron.
$h_1^{\perp g}$ being  a time-reversal even (T-even) TMD,  initial/final state interactions are not
necessary for the 
presence of 
$h_1^{\perp g}$. The gluon-gluon correlator \cite{Mulders:2000sh} of spin-$\frac12$ unpolarized proton is parametrized 
in terms of unpolarized
density of gluons, $f_1^g$ and $h_1^{\perp g}$. These are the only two TMDs of unpolarized proton which describe
the transverse dynamics of gluons. Without prior understanding of these functions, it is not
feasible to have theoretical interpretation of the physical quantities that are obtained experimentally. However, 
the studies on these functions
are still sparse. Hence, the determination of these functions must be of prime importance. \par
Though, the experimental investigations on the quantification of $h_1^{\perp g}$ are not carried out so far,
a model independent theoretical upper bound is given in
Ref. \cite{Boer:2010zf}. However, a lot of theoretical works have suggested  to 
probe $h_1^{\perp g}$ in several ways. Heavy quark pair or dijet production in SIDIS \cite{Pisano:2013cya},
diphoton pair \cite{Qiu:2011ai} and $\Upsilon(1\text{S}) +$jet \cite{Dunnen:2014eta} production in $pp$ collision have 
been suggested to probe $h_1^{\perp g}$. In these processes it has been shown that $h_1^{\perp g}$ can be
probed  by measuring  azimuthal asymmetries. Moreover, the participation of two linearly polarized gluons in
the scattering process results in a term  in the  cross section which is independent of azimuthal angle
\cite{Boer:2011kf}.  For instance, Higgs boson and quarkonium productions do not require    any angular analysis to extract 
$h_1^{\perp g}$. The Higgs $p_T$ distribution can be modified by taking linearly polarized gluons into account in unpolarized
$pp$ collision at LHC \cite{Echevarria:2015uaa,Boer:2011kf,Boer:2013fca}. Whether the Higgs is a scalar or pseudo scalar can be 
understood through the modified $p_T$ spectrum. It is also suggested that Higgs+jet production in $pp$ 
collision at LHC \cite{Boer:2014lka} is helpful to probe $h_1^{\perp g}$. Quarkonium production (even charge conjugation) has been studied in 
non-relativistic Quantum Chromodynamics (NRQCD) version of color singlet model (CSM) 
to investigate $h_1^{\perp g}$ \cite{Boer:2012bt}.\par
$J/\psi$ and $\Upsilon(1\text{S})$ bound states have been of   interest, not only because it is possible to measure the 
production  experimentally but also it gives information on the strong interaction responsible for
hadronization.  In the earlier work \cite{Mukherjee:2015smo}, we proposed $J/\psi$ and $\Upsilon(1\text{S})$ production 
in $pp$ collision to explore the effect of linearly polarized gluons using color evaporation model (CEM). The 
present article  discusses about probing $f_1^{ g}$ and $h_1^{\perp g}$ in quarkonium ($J/\psi$
and  $\Upsilon(1\text{S})$) production through  $gg$ fusion channel at leading order (LO)   in $pp$ collision using
color octet model (COM) in TMD factorization framework.\par
CSM, CEM and COM are three notable models for 
quarkonium production which are successful at different energies.  More
recently, in Ref. \cite{Ma:2015sia,Ma:2014mri}, color glass condensate 
model (CGS) and NRQCD NLO framework have been used to explain the full $p_T$ spectrum of $J/\psi$  in proton-proton and
proton-nucleus collision at RHIC.  All these models are based on a common
assumption : that the production rate is factorized into a product of short and long distance factors. The
production of $Q\bar{Q}$ pair calculated through perturbation theory, depends on specific process like $ep$ 
or $pp$. The produced $Q\bar{Q}$ pair transmutes into physical color singlet meson by radiating soft gluons.
This process can not be calculated using perturbative QCD tools. Moreover, there are two scales 
\cite{Amundson:1995em,Amundson:1996qr}
involved in the quarkonium production. The $Q\bar{Q}$ pair 
production takes place within a short time of order $1/M_{Q\bar{Q}}$, where $M_{Q\bar{Q}}$ is the mass of 
quarkonium. The binding process happens at a long time of order $1/\varLambda_{QCD}$ 
\cite{Amundson:1995em,Amundson:1996qr}.\par
CSM model was developed by  E. L. Berger and D. Jones \cite{Berger:1980ni,Baier:1981uk,Baier:1981zz} after the 
discovery of $J/\psi$ \cite{Augustin:1974xw,Aubert:1974js}. In CSM, the 
heavy quark pair is produced in color singlet state with certain spin (S) and orbital angular quantum number
(L) and later it evolves into  quarkonium, nonrelativistic bound state. The long distance factor (wave function
or its derivative) contains the nonperturbative information  of quarkonium. Generally, these objects are 
obtained from potential models or experimental data. CSM was able to predict the quarkonium production at low
energy \cite{Schuler:1994hy} but at high energy it requires large corrections in $\alpha_s$ 
\cite{Artoisenet:2007xi,Campbell:2007ws}. In CEM model, as introduced by
F.Halzen, Matsuda \cite{Halzen:1977rs,Halzen:1977im} and Fritsch \cite{Fritzsch:1977ay}, the production rate is
product of a long distance factor and the cross section of heavy quark pair integrated over the invariant mass of the $Q\bar{Q}$ pair
below the  threshold mass. In CEM, it is assumed that the probability of producing the quarkonium state is independent of
color and spin quantum number of initial heavy quark pair. It implies that the color of initial $Q\bar Q$ pair  does not
play any role in hadronization process. It was successful in describing $p_T$ distribution of $J/\psi$, $\psi(2S)$
and $\chi$ at  $\sqrt{s}=1.8$ TeV (CDF) \cite{Amundson:1995em,Amundson:1996qr}.\par
Bodwin, Braaten and Lepage \cite{Bodwin:1992ye} proposed COM based on NRQCD effective field theory. It is 
assumed  in COM that the heavy quark relative 
momentum ($M_Q\upsilon$) is much less than the mass ($M_Q$) of heavy quark, where $\upsilon$ is the relative velocity
of heavy quark in the rest frame of quarkonium. The ignored relativistic corrections in CSM are included in COM. The cross section of quarkonium is expressed as an infinite series in the limit $\upsilon \ll 1$.
Each term in the series is the
product of $Q\bar{Q}$ pair cross section in a definite  state $n=\leftidx{^{2S+1}}{L}{^{[a]}_J}$
and long distance matrix elements (LDME). Here, $J$, $L$ and $S$ are total angular momentum, orbital angular
momentum and spin quantum numbers respectively. $a$ is the color multiplicity bearing a value of 1 for color singlet and 
8 for color octet state. In this model, both the color singlet and color octet states are included in the  
production rate of quarkonium. The infinite series can be truncated for practical purpose.
LDME describes the transition probability to form the quarkonium
state from the heavy quark pair and is extracted from experimental data.
The COM  gives a good description of $J/\psi$  at  RHIC  energy \cite{Cooper:2004qe}. The present paper employs the 
NRQCD based model to study the TMDs and their evolution. Moreover, we compare the $J/\psi$
and $\Upsilon(1\text{S})$ 
production obtained in COM with earlier work which implemented CEM \cite{Mukherjee:2015smo}. The paper is organized 
as the following. Quarkonium production is presented in Sec.\ref{sec2} and Sec.\ref{sec3} describes the 
TMD evolution formalism. Numerical results and conclusions are presented in Sec.\ref{sec4} and Sec.\ref{sec5}
respectively.

\section{Quarkonium production}\label{sec2}
NRQCD formalism provides a theoretical framework to calculate the cross section of bound states.
As per NRQCD, the differential cross section of any quarkonium state is factorized as the following
\cite{Fleming:1995id,Bodwin:1994jh}
\be\label{e1}
d\sigma^{J/\psi(\Upsilon(1\text{S}))}=\sum_n d{\hat{\sigma}}[ab\rightarrow Q\bar{Q}(n)]
\langle 0\mid \mathcal{O}^{J/\psi(\Upsilon(1\text{S}))}_n\mid 0\rangle,
\ee
where $d{\hat\sigma}$ is the partonic level differential cross section of  $Q\bar{Q}$ pair calculated in 
order $\alpha_{s}$. The color, spin and angular momentum quantum numbers are collectively denoted with $n$.
Here, $a$ and $b$ can be gluons and quarks. The $Q$ ($\bar{Q}$) is a heavy quark (heavy anti-quark). The LDME, 
$\langle 0\mid \mathcal{O}^{J/\psi(\Upsilon(1\text{S}))}_n\mid 0\rangle$, contains the nonperturbative
effects like hadronization of 
$Q\bar{Q}$ pair which is defined as a four fermion operator \cite{Fleming:1995id}.
We consider the unpolarized $pp$ collision process for the charmonium and bottomonium production
\be
P(p_A)+P(p_B)\rightarrow J/\psi~\mathrm{or}~\Upsilon(1\text{S})(p)+~X,
\ee
the four momenta are given within round brackets. We consider a frame where the colliding protons 
are moving along the $+\hat{z}$ axis and $-\hat{z}$ axis. The four momenta in the center of mass frame are
$P^{\mu}_A=\frac{\sqrt{s}}{2}(1,0,0,1)$ and $P_B^\mu=\frac{\sqrt{s}}{2}(1,0,0,-1)$. The leading channel is 
gluon-gluon collision since proton is rich of gluons at high energy. Therefore
we consider the leading order (LO) gluon fusion subprocess for the quarkonium production, i.e., 
$gg\rightarrow Q\bar{Q}[\leftidx{^{2s+1}}{L}{^{(a)}_J}]$.
In line with Ref. \cite{Boer:2012bt,Boer:2007nd}, we assume that the QCD factorization theorem can be applicable at high 
energy to separate out the short and long distance effects systematically.  Using TMD factorization and NRQCD
formalism, the differential cross section of the quarkonium is given by 
\begin{equation}\label{cross1}
 \begin{aligned} 
 {d\sigma}={}&\int dx_{a} dx_{b}  d^{2}{\bf k}_{\perp a} d^2{\bf k}_{\perp b}
 \Phi^{\mu\nu}_g(x_{a},{\bf k}_{\perp a})\Phi_{g\mu\nu}(x_{b},{\bf k}_{\perp b})
 {d\sigma^{J/\psi(\Upsilon(1\text{S}))}},
\end{aligned}
\end{equation}
where $\Phi^{\mu\nu}_g(x_{a},{\bf k}_{\perp a})$ is the gluon correlator of the spin-$\frac{1}{2}$ 
unpolarized proton which is parametrized in terms of leading twist traditional TMDs \cite{Mulders:2000sh} as
following
\be \label{corre}
\Phi^{\mu\nu}_g(x,{\bf k}_{\perp})=-\frac{1}{2x}\left\{g^{\mu\nu}_Tf^g_1(x,{\bf k}_{\perp}^2)-\left(\frac{k^{\mu}_{\perp}k^{\nu}_{\perp}}
{M^2_h}+g^{\mu\nu}_T\frac{{\bf k}^2_{\perp}}{2M_h^2}\right)h^{\perp g}_1(x,{\bf k}_{\perp}^2)\right\}.
\ee
Here, $x$ and $k_\perp$ represent the longitudinal momentum fraction and transverse momentum of the  gluon 
respectively. $M_h$ is the mass of the proton. $f^g_1(x,{\bf k}_{\perp}^2)$ describes the density of unpolarized
gluons and linearly polarized gluon distribution is represented by $h^{\perp g}_1(x,{\bf k}_{\perp}^2)$ inside an 
unpolarized proton.\par
The partonic differential cross section is given by \cite{Fleming:1995id}
\begin{equation}\label{partonic}
 \begin{aligned}
d\sigma^{J/\psi(\Upsilon(1\text{S}))}
={}&\frac{5\alpha_s^2\pi^3}{96m_Q^3}\frac12d^2{\bf p}_{T}dy \delta^4(p_a+p_b-p)
\Big\{\langle 0\mid \mathcal{O}_8^{J/\psi(\Upsilon(1\text{S}))}(\leftidx{^1}{S}{_0})\mid 0\rangle\\
&+\frac{3}{m_Q^2}\langle 0\mid \mathcal{O}_8^{J/\psi(\Upsilon(1\text{S}))}(\leftidx{^3}{P}{_0})\mid 0\rangle 
 +\frac{4}{5m_Q^2}\langle 0\mid \mathcal{O}_8^{J/\psi(\Upsilon(1\text{S}))}(\leftidx{^3}{P}{_2})\mid 0\rangle\Big\}\\
={}&C_nd^2{\bf p}_{T}dy \delta^4(p_a+p_b-p),
\end{aligned}
\end{equation}
where, the four momentum vectors of the incoming gluons in center of mass frame are denoted with 
$ p_{a}=x_{a}\frac{\sqrt{s}}{2}\left(1+\frac{k^2_{\perp a}}{x^2_{a}s},\frac{2{\bf k}_{\perp a}}{x_{a}\sqrt{s}},
 1-\frac{k^2_{\perp a}}{x^2_{a}s}\right)$ and  
$ p_{b}=x_{b}\frac{\sqrt{s}}{2}\left(1+\frac{k^2_{\perp b}}{x^2_{b}s},\frac{2{\bf k}_{\perp b}}{x_{b}\sqrt{s}},
 -1+\frac{k^2_{\perp b}}{x^2_{b}s}\right)$ \cite{Anselmino:2002pd}.
The $Q\bar Q$ pair four momentum is $p=(p_0,{\bf p}_T,p_L)$.
Here, $m_Q$ is mass of heavy quark. ${\bf p}_{T}$ and $y$ are the transverse momentum and rapidity of the 
quarkonium respectively. $C_n$ is defined as the following
\begin{equation}
 \begin{aligned}
 C_n={}&\frac{5\alpha_s^2\pi^3}{96m_Q^3}\frac12\Big\{\langle 0\mid \mathcal{O}_8^{J/\psi(\Upsilon(1\text{S}))}
 (\leftidx{^1}{S}{_0})\mid 0\rangle
+\frac{3}{m_Q^2}\langle 0\mid \mathcal{O}_8^{J/\psi(\Upsilon(1\text{S}))}(\leftidx{^3}{P}{_0})\mid 0\rangle \\
 &+\frac{4}{5m_Q^2}\langle 0\mid \mathcal{O}_8^{J/\psi(\Upsilon(1\text{S}))}(\leftidx{^3}{P}{_2})\mid 0\rangle\Big\}.
 \end{aligned}
\end{equation}
As per Ref. \cite{Fleming:1995id,Cooper:2004qe}, the color octet states $\leftidx{^{1}}{S}{_0}$, $\leftidx{^{3}}{P}{_0}$ and
$\leftidx{^{3}}{P}{_2}$
contribution is dominant  for charmonium and bottomonium production  in
$gg\rightarrow Q\bar{Q}[\leftidx{^{2S+1}}{L}{^{(a)}_J}]$ subprocess.
The scattering amplitudes of $\leftidx{^{3}}{S}{_1}$ and $\leftidx{^{3}}{P}{_1}$ states vanish when two initial
scattering gluons are on-shell gluons \cite{Fleming:1995id,Cho:1995ce,Cho:1995vh}. We consider two 
sets (\textquotedblleft Set-I\textquotedblright and  \textquotedblleft Set-II\textquotedblright) 
of LDMEs for quarkonium production. 
In Set-I, the numerical values of LDME for $J/\psi$ \cite{Ma:2014mri,Chao:2012iv} and 
$\Upsilon(1\text{S})$ \cite{Sharma:2012dy} are  extracted by 
fitting  $J/\psi$ data at Tevatron and $\Upsilon(1\text{S})$ production at LHC in NLO 
collinear factorization using NRQCD framework. For Set-II, LDMEs are taken from Ref. 
\cite{Gong:2012ug,Brambilla:2014jmp} and \cite{Gong:2013qka} for $J/\psi$  and 
$\Upsilon(1\text{S})$ respectively. Feeddown contributions from $\chi_c$ and $\psi(2s)$ were 
included for extracting LDMEs of $J/\psi$ in Set-II unlike in Set-I.
LDME numerical values are given tabular form in \tablename{~\ref{table1}}.
\begin{table}[h!]
  \centering
  \caption{Numerical values of LDME.}
  \label{table1}
   \begin{tabular}{ccc}
    \toprule
  \hline
  \hline
 $ \langle 0\mid \mathcal{O}_8^{J/\psi(\Upsilon(1\text{S}))}(\leftidx{^{2S+1}}{L}{_J})\mid 0\rangle$ 
 &~~~Set-I \cite{Ma:2014mri,Chao:2012iv,Sharma:2012dy} &~~~Set-II 
\cite{Gong:2012ug,Brambilla:2014jmp,Gong:2013qka} \\
\hline
\midrule
  $\langle 0\mid \mathcal{O}_8^{J/\psi}(\leftidx{^1}{S}{_0})\mid 0\rangle$/GeV$^3$ &~~0.089  & 
~~0.097 \\
  $\langle 0\mid \mathcal{O}_8^{J/\psi}(\leftidx{^3}{P}{_0})\mid 0\rangle$/GeV$^5$ & ~~0.0126  
&~~ $-0.0214$ \\
$\langle 0\mid \mathcal{O}_8^{J/\psi}(\leftidx{^3}{P}{_2})\mid 0\rangle$/GeV$^5$ &  ~~0.063  
&~~$-0.107$ \\
$\langle 0\mid \mathcal{O}_8^{\Upsilon(1\text{S})}(\leftidx{^1}{S}{_0})\mid 0\rangle$/GeV$^3$ & 
~~0.0121  &~~ 0.111 \\
$\langle 0\mid \mathcal{O}_8^{\Upsilon(1\text{S})}(\leftidx{^3}{P}{_0})\mid 0\rangle$/GeV$^5$ & 
~~1.440  & ~~$-0.151$ \\
$\langle 0\mid \mathcal{O}_8^{\Upsilon(1\text{S})}(\leftidx{^3}{P}{_2})\mid 0\rangle$/GeV$^5$ & 
~~7.203   &~~$-0.755$ \\
  \hline 
  \hline
\bottomrule
  \end{tabular}
\end{table}
 The numerical value of LDME for $J=2$ state is 
obtained by using Eq.(6.6) from Ref. \cite{Bodwin:1994jh}.
The differential cross section in terms of TMDs is obtained by substituting Eq.\eqref{corre} and 
\eqref{partonic} in Eq.\eqref{cross1}
\begin{equation}
 \begin{aligned}
\frac{d^4\sigma}{dyd^2{\bf p}_{T}}={} &\frac{C_n}{2}\int \frac{dx_{a}}{x_a}\frac{dx_{b}}{x_b}d^{2}{\bf k}_{\perp a} d^2{\bf k}_{\perp b}
\delta^4(p_{a}+p_{b}-p) \\
&\times \left[f_1^g(x_{a},{\bf k}_{\perp a}^2)f_1^g(x_{b},{\bf k}_{\perp b}^2) 
 +wh_1^{\perp g}(x_{a},{\bf k}_{\perp a}^2)h_1^{\perp g}(x_{b},{\bf k}_{\perp b}^2)\right]
 \end{aligned}
\end{equation}
where $w$ is the transverse momentum weight factor 
\be
w=\frac{1}{2M_h^4}\left[({\bf k}_{\perp a}.{\bf k}_{\perp b})^2-\frac{1}{2}
{\bf k}_{\perp a}^2{\bf k}_{\perp b}^2\right].
\ee
The four momentum conservation delta function can be written as \cite{Anselmino:2002pd}
\be
 \delta^4(p_{a}+p_{b}-q)&=&\delta(E_{a}+E_{b}-p_0)\delta(p_{za}+p_{zb}-p_L)
  \delta^2({\bf k}_{\perp a}+{\bf k}_{\perp b}-{\bf p}_T)\\
  &=&\frac{2}{s}\delta\left(x_{a}-\frac{Me^{y}}{\sqrt{s}} \right) \label{delta}
  \delta\left(x_{b}-\frac{Me^{-y}}{\sqrt{s}} \right)
  \delta^2({\bf k}_{\perp a}+{\bf k}_{\perp b}-{\bf p}_T),
\ee
where, $M$ is the mass of quarkonium. After  integrating with respect to $x_{a}$ and $x_{b}$, the two $\delta$ 
functions contained in the above equation gives
\be\label{xab}
x_{a,b}=\frac{M}{\sqrt{s}}e^{\pm y}.
\ee
We can also eliminate ${\bf k}_{\perp b}$ by integrating and we get
\be
\frac{d\sigma^{ff+hh}}{dyd^2{\bf p}_T}=\frac{d\sigma^{ff}}{dyd^2{\bf p}_T}+\frac{d\sigma^{hh}}{dyd^2{\bf p}_T},
\ee
where
\begin{equation}
\begin{aligned}
\frac{d\sigma^{ff}}{dyd^2{\bf p}_T}={} &\frac{C_n}{s} \int d^2{\bf k}_{\perp a}
f_1^g(x_{a},{\bf k}_{\perp a}^2)
 f_1^g(x_{b},({\bf p}_{T}-{\bf k}_{\perp a})^2)
\end{aligned}
\end{equation}
and
\begin{equation}
\begin{aligned}
\frac{d\sigma^{hh}}{dyd^2{\bf p}_T}={} &\frac{C_n}{s}\frac{1}{2M_h^4} \int 
d^2{\bf k}_{\perp a} 
\left[\left({\bf k}_{\perp a}.({\bf p}_{T}-{\bf k}_{\perp a})\right)^2-\frac{1}{2}
{\bf k}_{\perp a}^2({\bf p}_{T}-{\bf k}_{\perp a})^2 \right]\\
&\times h_1^{\perp g}(x_{a},{\bf k}_{\perp a}^2)h_1^{\perp g}(x_{b},({\bf p}_{T}-{\bf k}_{\perp a})^2)
\end{aligned}
\end{equation}
\section{TMD Evolution}\label{sec3}
In this section, Dokshitzer-Gribov-Lipatov-Altarelli-Parisi (DGLAP) and TMD evolutions are discussed. 
Generally, we assume that the unpolarized gluon TMDs exhibit Gaussian distribution. The widely used  Gaussian 
parametrization of TMDs is given by
\be \label{unp}
 f_1^{g}(x,{\bf k}^2_{\perp })=f_1^{g}(x,Q^2)\frac{1}{\pi \langle k^2_{\perp }\rangle}
 e^{-{\bf k}^2_{\perp }/\langle k^2_{\perp }\rangle}.
\ee
Here, we factorized the TMD PDF into $x$ and $k_\perp$ dependencies. $f_1^{g}(x,Q^2)$ is the collinear PDF measured at the probing
scale $Q^2=M^2$ and scale evolution in  the $k_\perp$ dependent term is not taken into consideration. This is called 
the DGLAP evolution approach.
The Gaussian form of $h_1^{\perp g}$ is  the following \cite{Boer:2012bt}
\be\label{hg}
h_1^{\perp g}(x,{\bf k}^2_{\perp })=\frac{M^2_hf_1^g(x,Q^2)}{\pi\langle k^2_{\perp }\rangle^2}\frac{2(1-r)}{r}e^{1-
 {\bf k}^2_{\perp }\frac{1}{r\langle k^2_{\perp }\rangle}},
\ee
where, $r$ ($0<r<1$) is the parameter. We take $\langle k^2_{\perp }\rangle=0.25$ GeV$^2$  and 
1 GeV$^2$ \cite{Boer:2012bt} and  $r=\frac13$  and  $\frac23$  \cite{Boer:2012bt} for numerical calculation. 
Model independent upper bound for $h_1^{\perp g}$  is given in \cite{Boer:2010zf} and is obeyed by the Eq.\eqref{hg} for all values of $k_\perp$
and $x$
\be
\frac{{\bf k}_{\perp}^2}{2M_h^2}|h^{\perp }_1(x,{\bf k}_{\perp }^2)|\leq f_1^g(x,{\bf k}^2_{\perp }).
\ee
\subsection{Model-I}
In model-I, an upper limit of transverse momentum integration is not chosen.
The Gaussian form of the unpolarized and linearly polarized TMDs allow us to integrate analytically and we get
\begin{equation}\label{m1ff}
\begin{aligned}
\frac{d^2\sigma^{ff}}{dyd p^2_T}={} &\frac{C_n\beta}{2s}e^{-\frac{\beta p_T^2}{2}}f_1^g(x_{a})f_1^g(x_{b}),
 \end{aligned}
\end{equation}
and
\begin{equation}\label{m1hh}
\begin{aligned}
\frac{d^2\sigma^{hh}}{dyd p^2_T}={} &\frac{C_n\beta (1-r)^2re^2}{4s}e^{-\frac{\beta p_T^2}{2r}}
\left[1-\frac{\beta p^2_T}{r}+\frac{\beta^2p_T^4}{8r^2}\right]f_1^g(x_{a})f_1^g(x_{b}).
 \end{aligned}
\end{equation}
where $\beta=\frac{1}{\langle k^2_{\perp a}\rangle}=\frac{1}{\langle(p_T- k_{\perp a})^2\rangle}$.
\subsection{Model II}
In model-II,  for Gaussian distribution functions the effective intrinsic motion of partons is limited to
$k_{\mathrm{max}}=\sqrt{\langle {k}^2_{\perp a}\rangle}$ \cite{Anselmino:2008sga}. The final expressions we have
\begin{equation}\label{m2ff}
\begin{aligned}
\frac{d^2\sigma^{ff}}{dyd p^2_T}={} &\frac{C_n\beta^2}{2s\pi^2}\int_0^{2\pi} d\phi_{p_T}
\int_0^{2\pi}d\phi_{k_{\perp a}}\int_0^{\sqrt{\langle {k}^2_{\perp a}\rangle}} k_{\perp a}dk_{\perp a} 
 e^{-\beta \Delta}f_1^g(x_{a})f_1^g(x_{b}),
 \end{aligned}
\end{equation}
and
\begin{equation}\label{m2hh}
\begin{aligned}
\frac{d^2\sigma^{hh}}{dyd p^2_T}={} &\frac{C_n\beta^4}{s\pi^2}\frac{(1-r)^2e^2}{r^2}
\int_0^{2\pi} d\phi_{p_T}
\int_0^{2\pi}d\phi_{k_{\perp a}}\int_0^{\sqrt{\langle {k}^2_{\perp a}\rangle}} k_{\perp a}dk_{\perp a} 
 \Big[ \frac{1}{2}k^4_{\perp a}
-\frac{1}{2}k^2_{\perp a}p^2_T\\
&-p_Tk^3_{\perp a} \cos(\phi_{k_{\perp a}}-\phi_{p_T})
+p^2_T k^2_{\perp a}\cos^2(\phi_{k_{\perp a}}-\phi_{p_T}) \Big] 
 e^{-\frac{\beta}{r} \Delta}f_1^g(x_{a})f_1^g(x_{b}).
\end{aligned}
\end{equation}
where  $\Delta=2k^2_{\perp a}+p^2_T-2p_Tk_{\perp a}\cos(\phi_{k_{\perp a}}-\phi_{p_T})$.
The azimuthal angle of gluon and quarkonium are $\phi_{k_{\perp a}}$ and  $\phi_{p_T}$ respectively. We have chosen 
$\langle k^2_{\perp a}\rangle=\langle k^2_{\perp }\rangle$ for numerical estimation of quarkonium production rate.
DGLAP evolution approach could not describe the  Z-boson high transverse momentum distribution  in DY
process at CDF \cite{Melis:2014pna}. Nevertheless, one has to consider TMD evolution approach to explain high $p_T$
data \cite{Melis:2014pna}.
In TMD evolution approach, we follow the formalism adopted in Ref. \cite{Mukherjee:2015smo} to 
study scale evolution of TMDs. 
The quarkonium differential cross section in $b_\perp$-space is derived by following Eq.(29) to Eq.(33) from 
Ref. \cite{Mukherjee:2015smo}
\begin{equation}\label{et4}
 \begin{aligned} 
 \frac{d\sigma}{dyd^2{\bf p}_T}={}&\frac{C_n}{s}\frac{1}{2\pi}
 \int_0^{\infty}b_{\perp} db_{\perp}J_0(p_Tb_{\perp})
 \Big\{ f_1^g(x_{a}, b_{\perp}^2)f_1^g(x_{b}, b_{\perp}^2)+h_1^{\perp g}(x_{a}, b_{\perp}^2)h_1^{\perp g}(x_{b},b_{\perp}^2)\Big\},
\end{aligned}
\end{equation}
where $J_0$ is the zeroth order Bessel function.
In TMD factorization theorem, spurious light cone divergences appear \cite{jcollins} which can be regularized by introducing auxiliary 
parameter $\zeta$. As a result, TMDs  depend on renormalization scale $\mu$ and $\zeta$. Using CS and RG
equations \cite{jcollins,Aybat:2011zv} one can evolve the TMDs from 
initial scale $Q_i=c/b_{\ast}(b_{\perp})=\zeta_0$ to  final scale
$Q_f=Q=\zeta$ \cite{Aybat:2011zv,Aybat:2011ta} 
\begin{eqnarray}{\label{pert}}
  f(x,b_{\perp},Q_f,\zeta)=f(x,b_\perp,Q_i,\zeta)R_{pert}\left(Q_f,Q_i,b_{\ast}\right)
  R_{NP}\left(Q_f,Q_i,b_{\perp}\right),
 \end{eqnarray}
where $R_{pert}$ is the perturbative part of the evolution kernel and can be calculated using perturbation
theory. $R_{NP}$ is the nonperturbative part of evolution kernel and TMDs. Initial scale of TMDs is chosen to
be $Q_i=c/b_{\ast}$, where $c=2e^{-\gamma_\epsilon}$ with $\gamma_\epsilon\approx0.577$. The initial scale $Q_i$
 becomes small when $b_\perp$ is large, as a result we enter in the nonperturbative regime \cite{jcollins}.
 The $b_{\ast}$ prescription
 is adopted to separate the evolution kernel nonperturbative part, where
 $b_{\ast}(b_{\perp})=\frac{b_{\perp}}{\sqrt{1+\left(\frac{b_{\perp}}{b_{\mathrm{max}}}\right)^2}}
 \approx b_{\mathrm{max}}$ when $b_{\perp}\rightarrow \infty$ and $b_{\ast}(b_{\perp})\approx
 b_{\perp}$ when $b_{\perp}\rightarrow 0$. $R_{NP}$ contains  nonperturbative information of 
 evolution kernel that cannot be calculated and need to be parametrized. The perturbative 
evolution kernel is given by \cite{Boer:2014tka}
 \be\label{sudakov}
 R_{pert}\left(Q_f,Q_i,b_{\ast}\right)=\mathrm{exp}\Big\{{-\int_{c/b_{\ast}}^{Q}\frac{d\mu}
 {\mu}\left(A\log\left(\frac{Q^2} {\mu^2}\right)+B\right)}\Big\},
 \ee
 where the anomalous dimensions are  denoted by $A$ an $B$ respectively and 
 these have perturbative  expansion as follows :
   $$A=\sum_{n=1}^{\infty}\left(\frac{\alpha_s(\mu)}{\pi}\right)^nA_n$$
 and $$B=\sum_{n=1}^{\infty}\left(\frac{\alpha_s(\mu)}{\pi}\right)^nB_n.$$ 
 $A_1=C_A$ and $B_1=-\frac{1}{2}(\frac{11}{3}C_A-\frac{2}{3}N_f)$ are the anomalous dimension coefficients of
 order in $\alpha_s$.
 The anomalous dimensions have been calculated up to 3-loop level \cite{Idilbi:2006dg}.
 The evolution kernel is the same for linearly polarized gluons since it  is
 independent of type of TMDs. As stated before, the non-perturbative part of
 the evolution kernel cannot be calculated, and a parametrized form  has to be chosen. 
 Here we use two nonperturbative factor parametrizations which are called \textquotedblleft 
AR\textquotedblright~ and \textquotedblleft BLNY\textquotedblright. The \textquotedblleft 
AR\textquotedblright~ nonperturbative Sudakov factor was cansidered by Aybat and 
Rogers \cite{Aybat:2011zv} and  is successful in
 describing the low energy SIDIS and DY data
 \be \label{rnp1}
R_{NP}(x,Q,b_\perp)=\mathrm{exp}\left\{-\left[\frac{g_2}{2}\log\frac{Q}{2Q_0}+\frac{g_1}{2}
\left(1+2g_3\log\frac { 10xx_0}{x_0+x}\right)
\right]b_{\perp}^2\right\}.
 \ee
 The \textquotedblleft BLNY\textquotedblright~Sudakov nonperturbative factor was used by Sun et 
al. \cite{Sun:2012vc} in quarkonium production and is given by

\be \label{rnp2}
R_{NP}(x,Q,b_\perp)=\mathrm{exp}\left\{-\left[\frac{g_2}{2}\log\frac{Q}{2Q_0}+\frac{g_1}{2}
+g_1g_3\log \left(10x\right)\right]b_{\perp}^2\right\}.
 \ee
 
\begin{table}[h!]
\centering
\begin{tabular}{| >{\centering\arraybackslash}m{2cm} | >{\centering\arraybackslash}m{2cm} 
|>{\centering\arraybackslash}m{2cm}|>{\centering\arraybackslash}m{2cm}|>{\centering\arraybackslash}m
{ 2cm}|>{\centering\arraybackslash}m{2.5cm}|>{\centering\arraybackslash}m{2cm}|}
  \hline
   ${R_{NP}}$ & $g_1$/GeV$^2$  & $g_2$/GeV$^2$ &$g_3$ &$Q_0$/GeV &b$_\mathrm{max}$/GeV$^{-1}$ & $x_0$ \\
  \hline 
 AR \cite{Aybat:2011zv}  & 0.201   & 0.184  &$ -0.129$ & 1.6 & 1.5 & 0.009  \\
  \hline
   BLNY \cite{Sun:2012vc}&0.03 & 0.87 & $-5.66$ & 1.6 &0.5 &\\
  \hline
  \end{tabular}  
\caption{\label{table2}Best fit parameters of nonperturbative Sudakov factor (${R_{NP}}$)}
\end{table}
 The numerical values of the best fit parameters are given in \tablename{~\ref{table2}}.
Though, $R_{NP}$ is $x$ dependent, we choose $x_a=x_b=0.09$ only for \textquotedblleft 
AR\textquotedblright~$R_{NP}$ as per 
Ref. \cite{Aybat:2011zv,Boer:2014tka} to write the  $R_{NP}$ in
the form of a Gaussian function. For \textquotedblleft 
BLNY\textquotedblright~$R_{NP}$, Eq.\eqref{xab} is used for $x_a$ and $x_b$. We choose 
Eq.\eqref{rnp1} and \eqref{rnp2} as the nonperturbative Sudakov factors for linearly
polarized gluon TMD PDF as well since no experimental data is available to extract the best fit parameters  
of $R_{NP}$ for $h^{\perp g}_1$.  In general,  TMDs are written as the convolution of coefficient function 
times the collinear PDF  
\be
f(x,b_\perp,Q_i,\zeta)=\sum_{i=g,q}\int_{x}^1\frac{d\hat{x}}{\hat{x}}C_{i/g}(x/\hat{x},b_{
\perp } , \alpha_s,Q_i,\zeta)
  f_{i/p}(\hat{x},c/b_{\ast})+\mathcal{O}(b_{\perp}\varLambda_{QCD}),
  \ee
where the coefficient function is dependent on the type of TMD and is independent of the process,  this  is 
calculated using perturbation theory.
The unpolarized and linearly polarized TMDs in terms of collinear PDFs at leading and first order in $\alpha_s$
are \cite{Boer:2014tka}
\be\label{et5}
  f_1^g(x,b_\perp,Q_i,\zeta)=f_{g/p}(x,c/b_{\ast})+\mathcal{O}(\alpha_s),
 \ee
 \be\label{et6}
 h_1^{\perp 
g}(x,b_\perp,Q_i,\zeta)=\frac{\alpha_s(c/b_{\ast})C_A}{\pi}\int_{x}^1\frac{d\hat{x}}{\hat{x}}
 \left(\frac{\hat{x}}{x}-1\right)
 f_{g/p}(\hat{x},c/b_{\ast})+\mathcal{O}(\alpha_s^2).
 \ee
 Using above equations one can rewrite Eq.\eqref{et4} as 
  \be \label{tmdevo}
\frac{d^2\sigma^{ff+hh}}{dydp^2_T}=\frac{d^2\sigma^{ff}}{dydp^2_T}+\frac{d^2\sigma^{hh}}{dydp^2_T},
\ee
where
 \begin{equation}\label{evoleq1}
\begin{aligned}
 \frac{d^2\sigma^{ff}}{dydp^2_T}={} &\frac{C_n}{2s}
 \int_0^{\infty}b_{\perp} db_{\perp}J_0(p_Tb_{\perp})
 f^g_1(x_a,c/b_{\ast})f_1^g(x_b,c/b_{\ast})
\mathrm{exp}\Bigg\{{-2\int_{c/b_{\ast}}^{Q}\frac{d\mu}{\mu}\left(A\log\left(\frac{Q^2}
 {\mu^2}\right)+B\right)}\Bigg\}\\
&~~~~~~\times R_{NP}(x_a,Q,b_\perp)R_{NP}(x_b,Q,b_\perp),
 \end{aligned}
\end{equation}
and 
\begin{equation}\label{evoleq2}
\begin{aligned}
 \frac{d^2\sigma^{hh}}{dydp^2_T}={} &\frac{C_n C_A^2}{2s\pi^2}
 \int_0^{\infty}b_{\perp} db_{\perp}J_0(p_Tb_{\perp})\alpha_s^2(c/b_{\ast})
 \int_{x_a}^1\frac{dx_1}{x_1}\left(\frac{x_1}{x_a}-1\right)f^g_1(x_1,c/b_{\ast})
\int_{x_b}^1\frac{dx_2}{x_2}\left(\frac{x_2}{x_b}-1\right)\\
&f_1^g(x_2,c/b_{\ast})
\mathrm{exp}\Bigg\{{-2\int_{c/b_{\ast}}^{Q}
\frac{d\mu}{\mu}\left(A\log\left(\frac{Q^2}
 {\mu^2}\right)+B\right)}\Bigg\}R_{NP}(x_a,Q,b_\perp)R_{NP}(x_b,Q,b_\perp).
 \end{aligned}
\end{equation}
 
\section{Numerical Results}\label{sec4}
MSTW2008 is used for numerical calculations \cite{Martin:2009iq}. Masses of $J/\psi$
and $\Upsilon(1\text{S})$ are taken as $M=3.096$ and 9.398 GeV respectively. The transverse momentum ($p_T$) and rapidity ($y$) distributions of $J/\psi$
and $\Upsilon(1\text{S})$ are estimated in unpolarized $pp$ collision at $\sqrt{s}=7$ TeV (LHCb), $\sqrt{s}=500$ GeV (RHIC) and 
$\sqrt{s}=115$ GeV (AFTER) in NRQCD formalism using TMD factorization. We have considered the color octet states  
in LO subprocess $gg\rightarrow Q\bar{Q}[\leftidx{^{2s+1}}{L}{^{8}_J}]$  for quarkonium production.
To obtain the cross section differential in $p_T$, the  integration of rapidity is chosen in the range of
$y\in[2.0,4.5]$, $y\in[-3.0,3.0]$ and  $y\in[-0.5,0.5]$ for LHCb, RHIC and AFTER respectively.
In general, the LDME ($\langle 0\mid \mathcal{O}^{J/\psi(\Upsilon(1\text{S}))}_n\mid 0\rangle$) in COM and $\rho$ in CEM depend on the 
mass of heavy quark ($m_Q$), scale $Q$, order of the calculation (LO, NLO) and PDFs $f(x,Q^2)$ \cite{Frawley:2008kk}.
Hence,  $m_c=1.5$ GeV and $m_b=4.88$ GeV in COM is chosen in line with Ref. \cite{Ma:2014mri} and \cite{Sharma:2012dy}
respectively. In CEM, $m_c=1.2$ GeV and $m_b=4.75$ GeV is considered \cite{Mukherjee:2015smo,smith} for charm and 
bottom quark masses.
\par
In all the figures, the conventions are the following.
\textquotedblleft ff\textquotedblright~represents the distribution of quarkonium and is obtained by 
taking  into account only the unpolarized gluon contribution in the scattering process. The distribution of quarkonium 
denoted with \textquotedblleft ff+hh\textquotedblright~ is obtained by considering both unpolarized and 
linearly polarized gluons in $pp$ collision. Two sets of LDMEs $i.e.,$ \textquotedblleft Set-I\textquotedblright~and 
\textquotedblleft Set-II\textquotedblright~are considered for color octet states
 which are given in \tablename{~\ref{table1}}. The $p_T$ and $y$ spectra of quarkonium
are estimated both in DGLAP and TMD evolution approach.\par
\begin{figure}[H]
\begin{minipage}{0.99\textwidth}
\includegraphics[width=7.5cm,height=6cm]{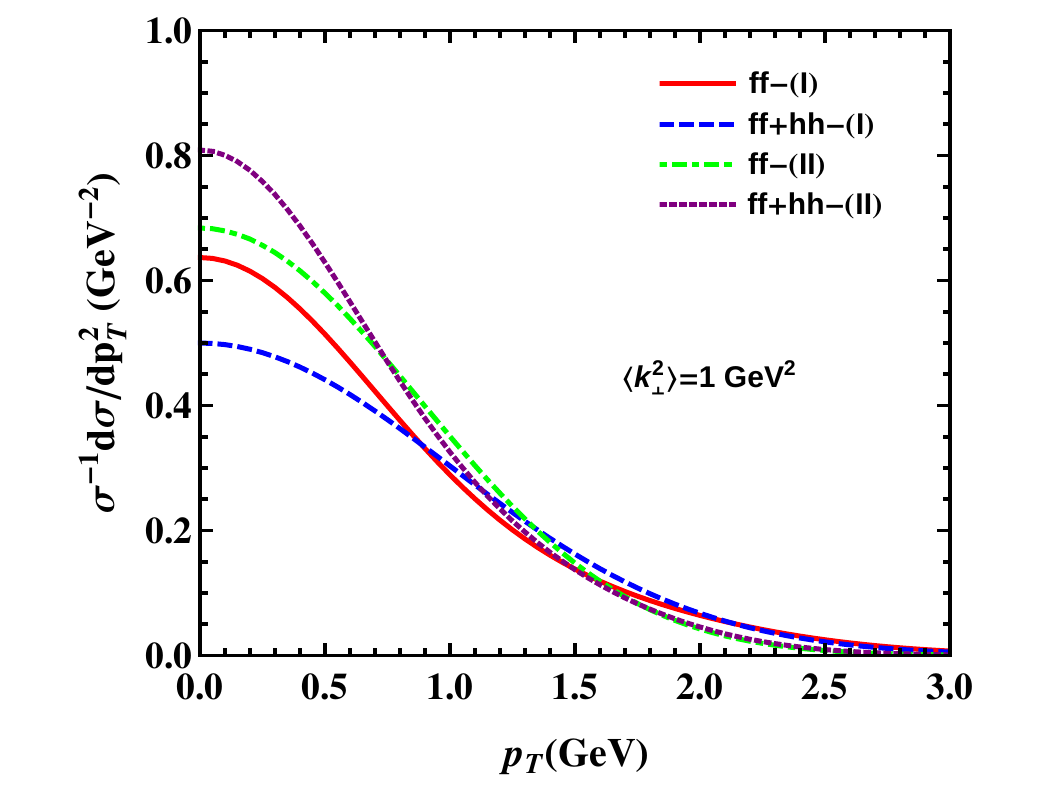}
\hspace{0.1cm}
\includegraphics[width=7.5cm,height=6cm]{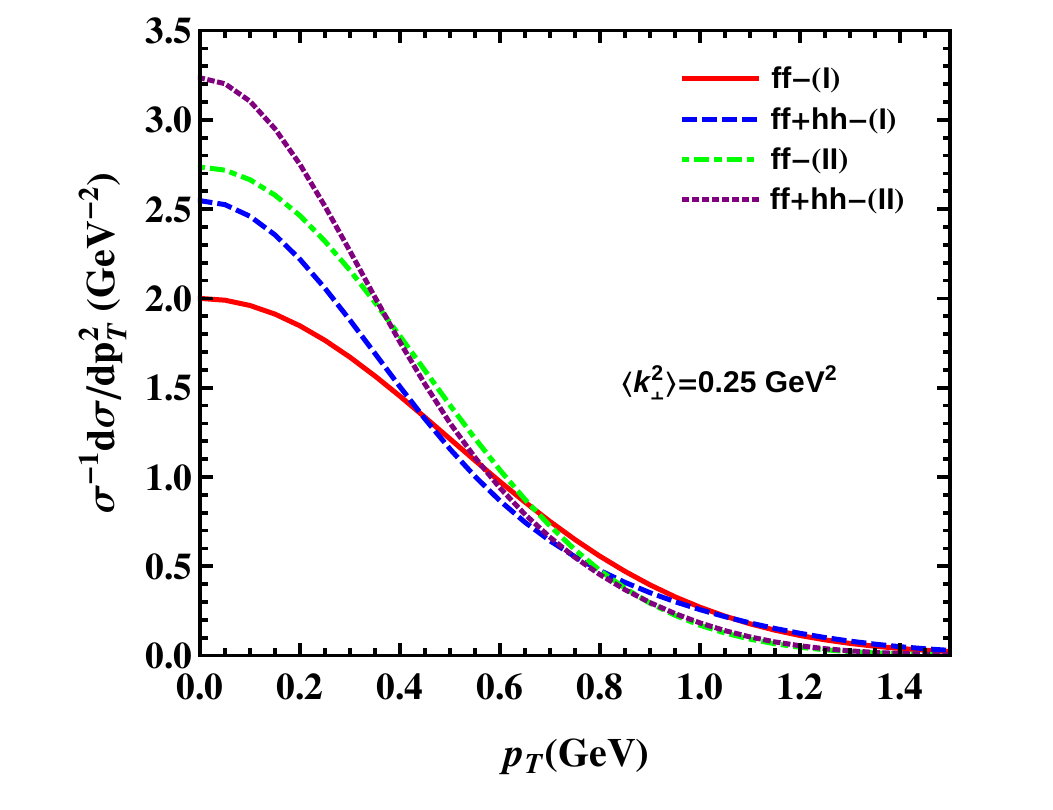}
\end{minipage}
\caption{\label{fig1}(color online) Differential cross section (normalized) of 
 $J/\psi$  and $\Upsilon(1\text{S})$ production in $pp\rightarrow J/\psi(\Upsilon(1\text{S}))+X$  
at LHCb ($\sqrt{s}=7$ TeV),  RHIC ($\sqrt{s}=500$ GeV) and AFTER ($\sqrt{s}=115$ GeV) energies
using  \textquotedblleft Set-I\textquotedblright~LDMEs in DGLAP evolution approach  for  
$r=\frac{2}{3}$.
The solid (ff-(I)) and dot dashed (ff-(II)) lines are obtained by considering
unpolarized gluons  in Model-I and Model-II respectively.
The dashed (ff+hh-(I)) and tiny dashed (ff+hh-(II)) lines are obtained by  taking 
into account unpolarized gluons plus linearly polarized gluons in Model-I and Model-II
respectively. See the text for ranges of rapidity integration.}
\end{figure}
\begin{figure}[H]
\begin{minipage}{0.99\textwidth}
\includegraphics[width=7.5cm,height=6cm,clip]{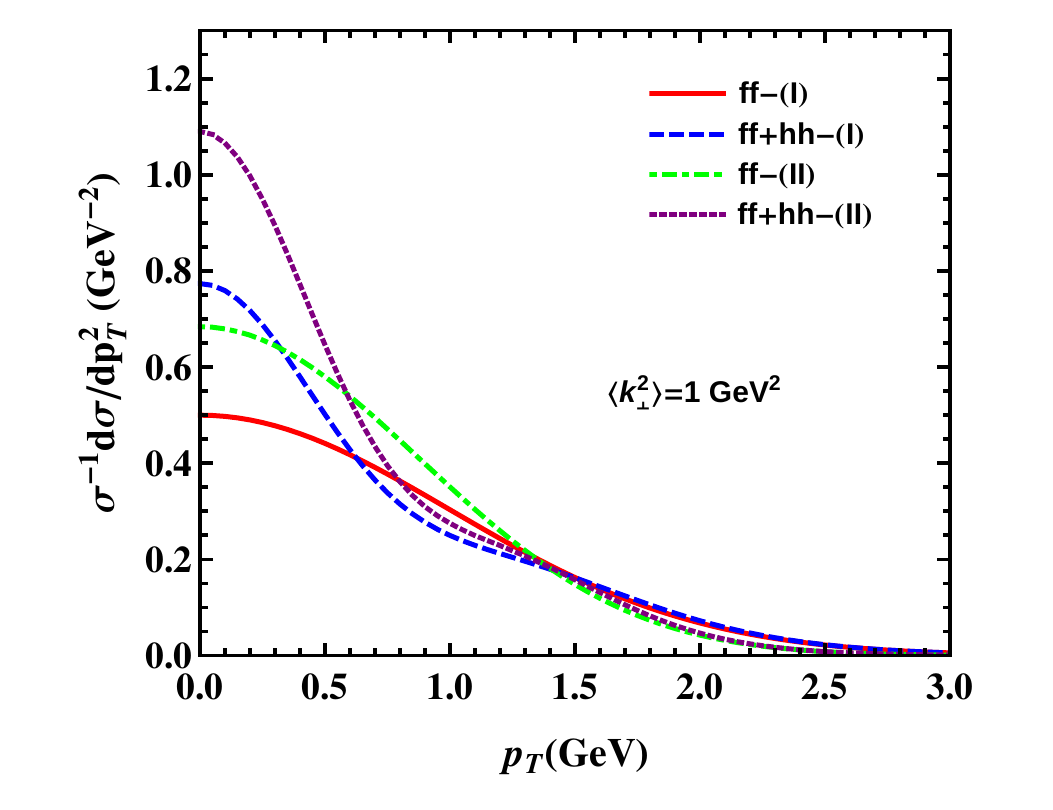}
\hspace{0.1cm}
\includegraphics[width=7.5cm,height=6cm,clip]{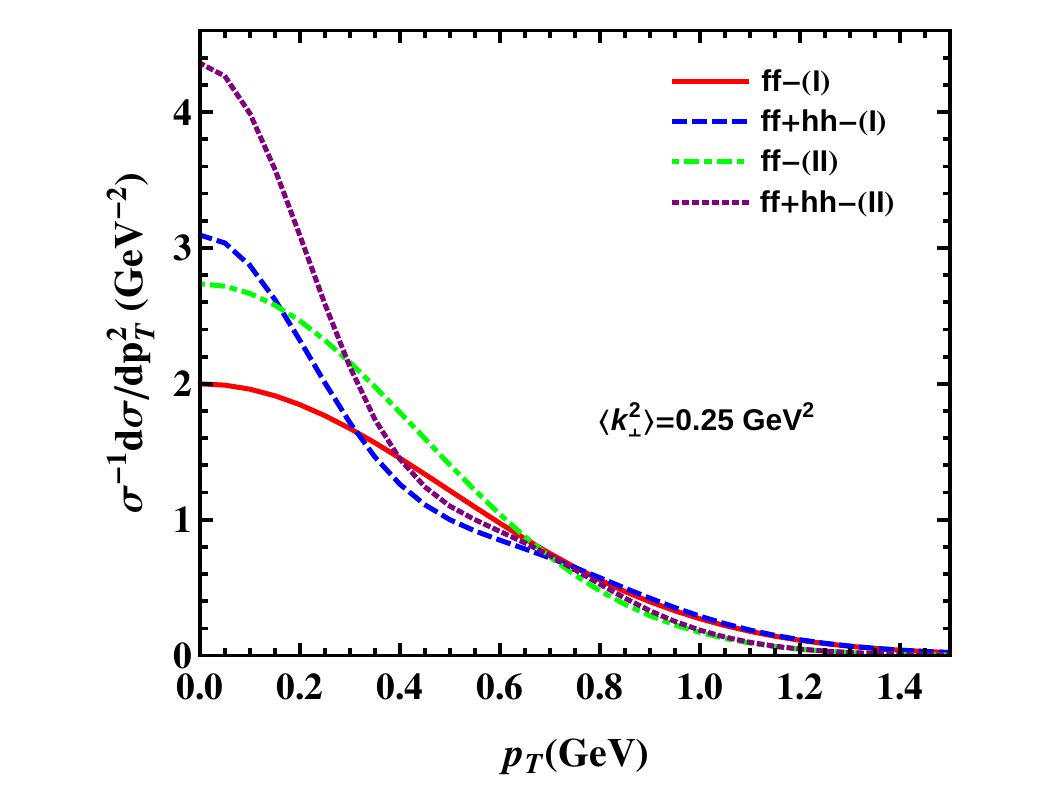}
\end{minipage}
\caption{\label{fig2}(color online) Same as in Fig. \ref{fig2} but for
$r=\frac{1}{3}$.}
\end{figure}
\begin{figure}[H]
\begin{minipage}[c]{0.99\textwidth}
\small{(a)}\includegraphics[width=7.5cm,height=6cm,clip]{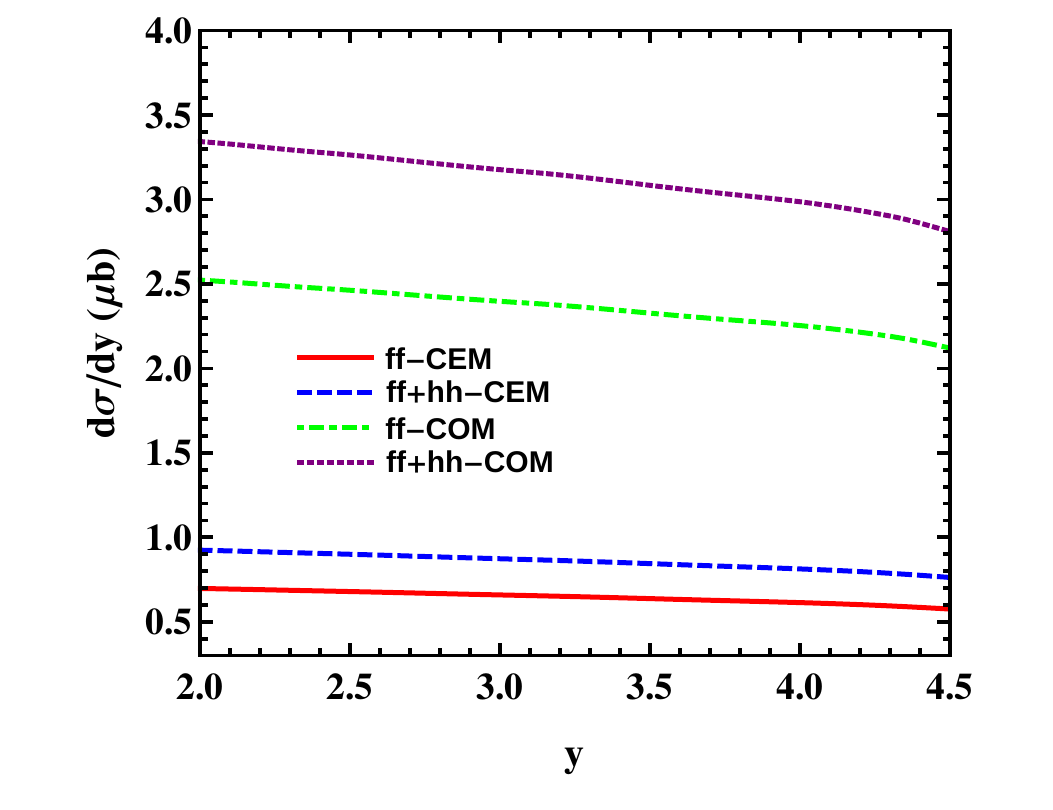}
\hspace{0.1cm}
\small{(b)}\includegraphics[width=7.5cm,height=6cm,clip]{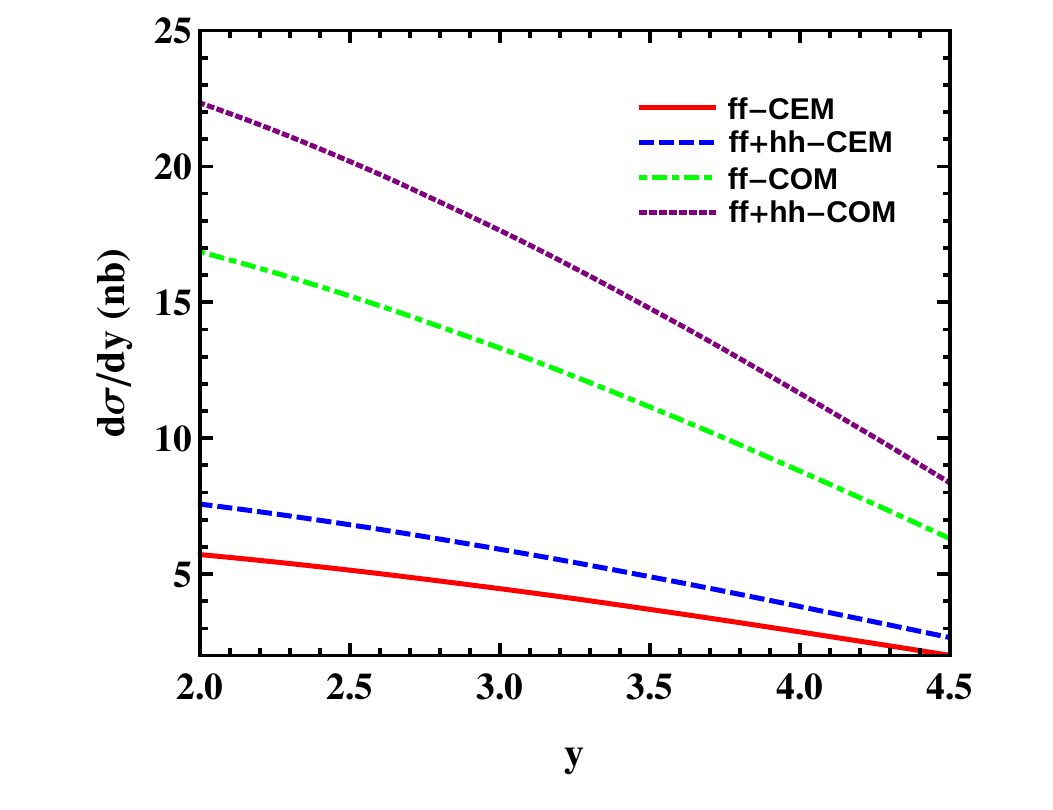}
\end{minipage}
\caption{\label{fig3}(color online). Rapidity ($y$) distribution of 
(a) $J/\psi$ (left panel) and  (b) $\Upsilon(1\text{S})$ (right panel) in $pp\rightarrow J/\psi(\Upsilon(1\text{S}))+X$
at LHCb ($\sqrt{s}=7$ TeV) energy and $p_T$  integration range is from 0 to 0.5 GeV 
using DGLAP evolution approach in Model-I for  $\langle k^2_{\perp}\rangle$=1 GeV$^2$ and
$r=\frac13$ in both CEM and COM. \textquotedblleft Set-I\textquotedblright~LDMEs are used in COM. 
The solid (ff-CEM) and dot dashed (ff-COM) lines are obtained by considering
unpolarized gluons  in CEM and COM respectively.
The dashed (ff+hh-CEM) and tiny dashed (ff+hh-COM) lines are obtained by  taking 
into account unpolarized gluons  plus linearly polarized gluons in CEM and COM
respectively.}
\end{figure}
\begin{figure}[H]
\begin{minipage}[c]{0.99\textwidth}
\small{(a)}\includegraphics[width=7.5cm,height=6cm,clip]{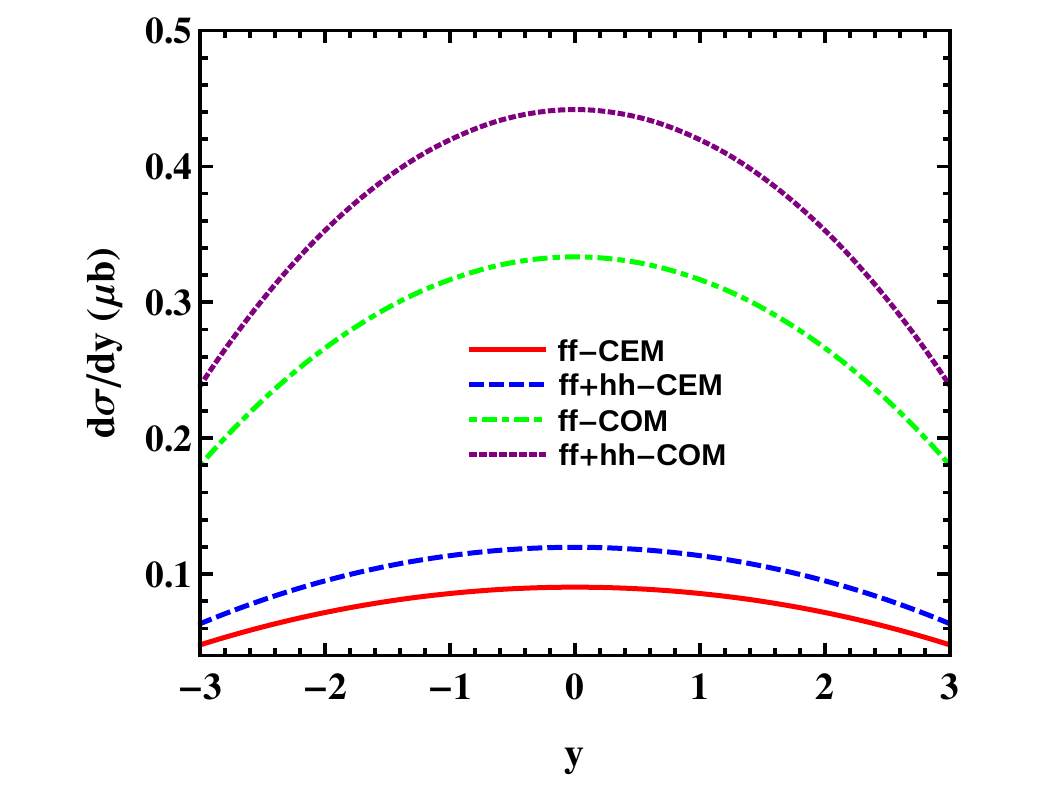}
\hspace{0.1cm}
\small{(b)}\includegraphics[width=7.5cm,height=6cm,clip]{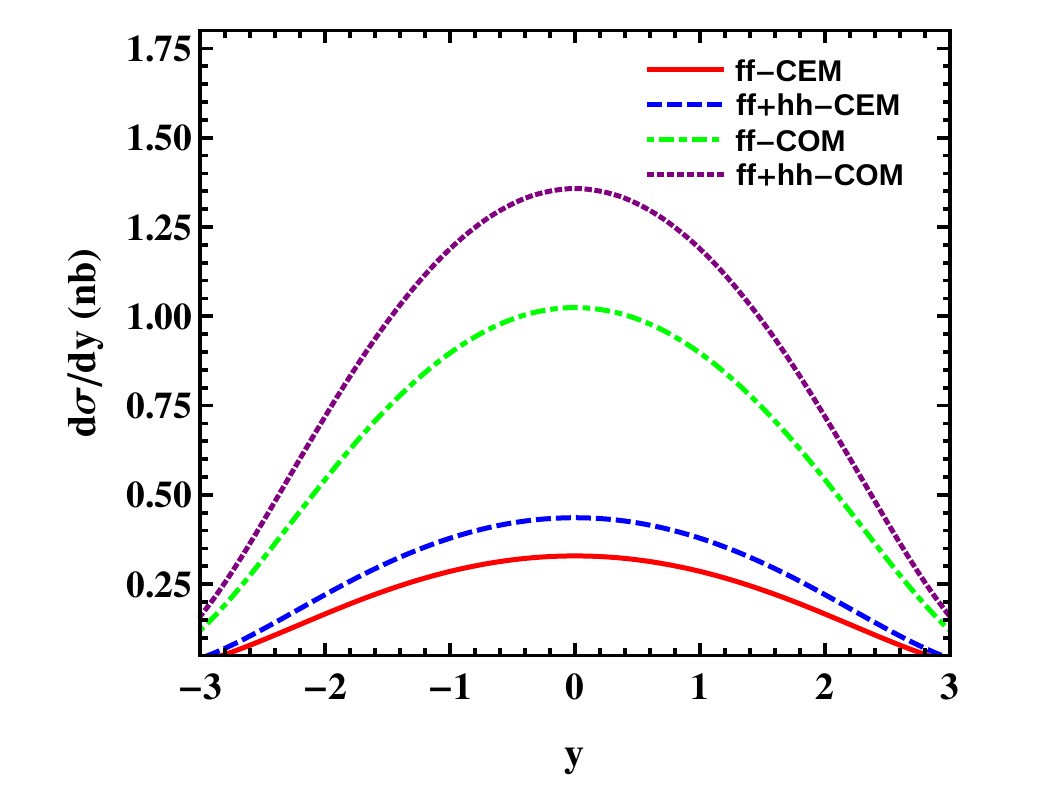}
\end{minipage}
\caption{\label{fig4}(color online). Rapidity ($y$) distribution of 
(a) $J/\psi$ (left panel) and  (b) $\Upsilon(1\text{S})$ (right panel) in $pp\rightarrow J/\psi(\Upsilon(1\text{S}))+X$
at RHIC ($\sqrt{s}=500$ GeV) energy and $p_T$  integration range is from 0 to 0.5 GeV 
using DGLAP evolution approach in Model-I for  $\langle k^2_{\perp}\rangle$=1 GeV$^2$ and $r=\frac13$ in 
both CEM and COM.  \textquotedblleft Set-I\textquotedblright~LDMEs are used in COM.  The convention 
in the figure for line styles is same as Fig. \ref{fig3}.}
\end{figure}
\begin{figure}[H]
\begin{minipage}[c]{0.99\textwidth}
\small{(a)}\includegraphics[width=7.5cm,height=6cm,clip]{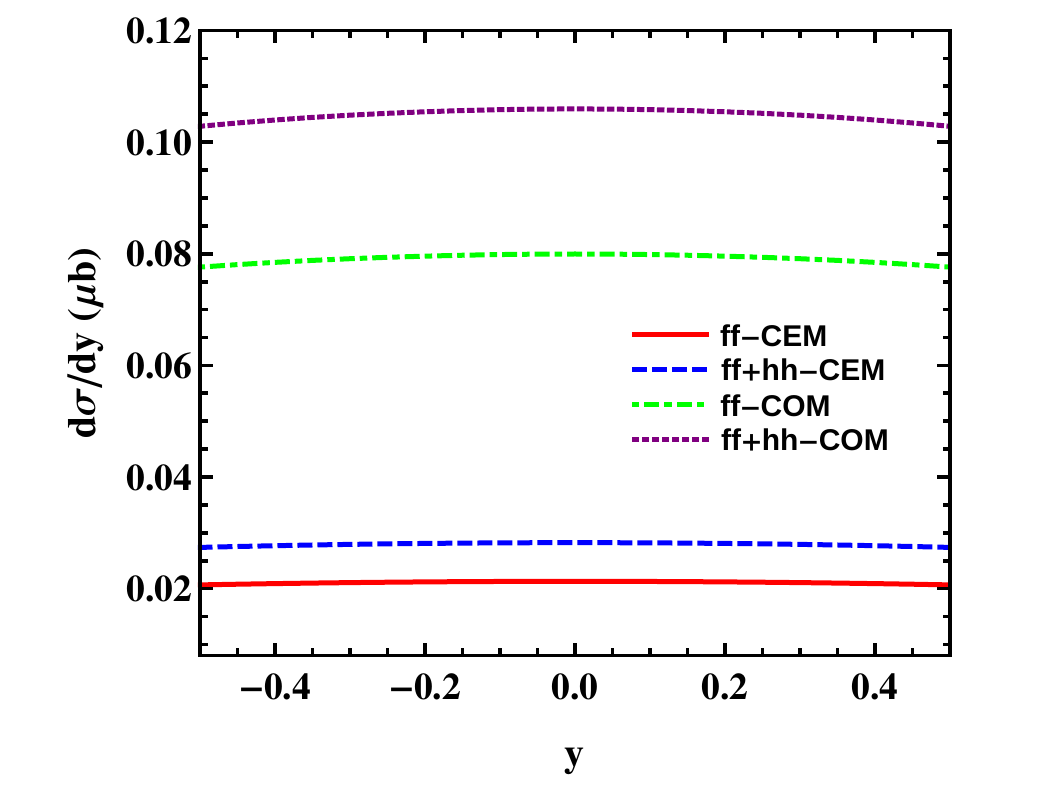}
\hspace{0.1cm}
\small{(b)}\includegraphics[width=7.5cm,height=6cm,clip]{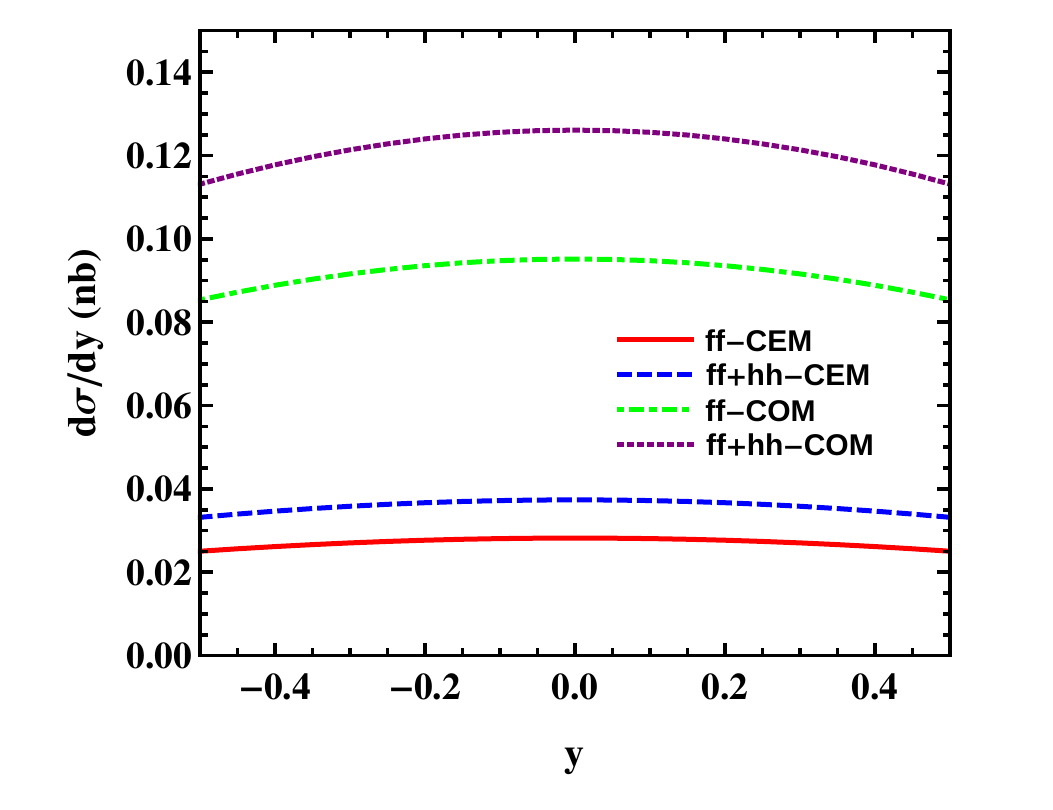}
\end{minipage}
\caption{\label{fig5}(color online). Rapidity ($y$) distribution of 
(a) $J/\psi$ (left panel) and  (b) $\Upsilon(1\text{S})$ (right panel) in $pp\rightarrow J/\psi(\Upsilon(1\text{S}))+X$
at AFTER ($\sqrt{s}=115$ GeV) energy and $p_T$  integration range is from 0 to 0.5 GeV 
using DGLAP evolution approach in Model-I for  $\langle k^2_{\perp}\rangle$=1 GeV$^2$ and $r=\frac13$ in 
both CEM and COM.  \textquotedblleft Set-I\textquotedblright~LDMEs are used in COM.  The convention 
in the figure for
 line styles is same as Fig. \ref{fig3}.}
\end{figure}
\begin{figure}[H]
\begin{minipage}[c]{0.99\textwidth}
\small{(a)}\includegraphics[width=7.5cm,height=6cm,clip]{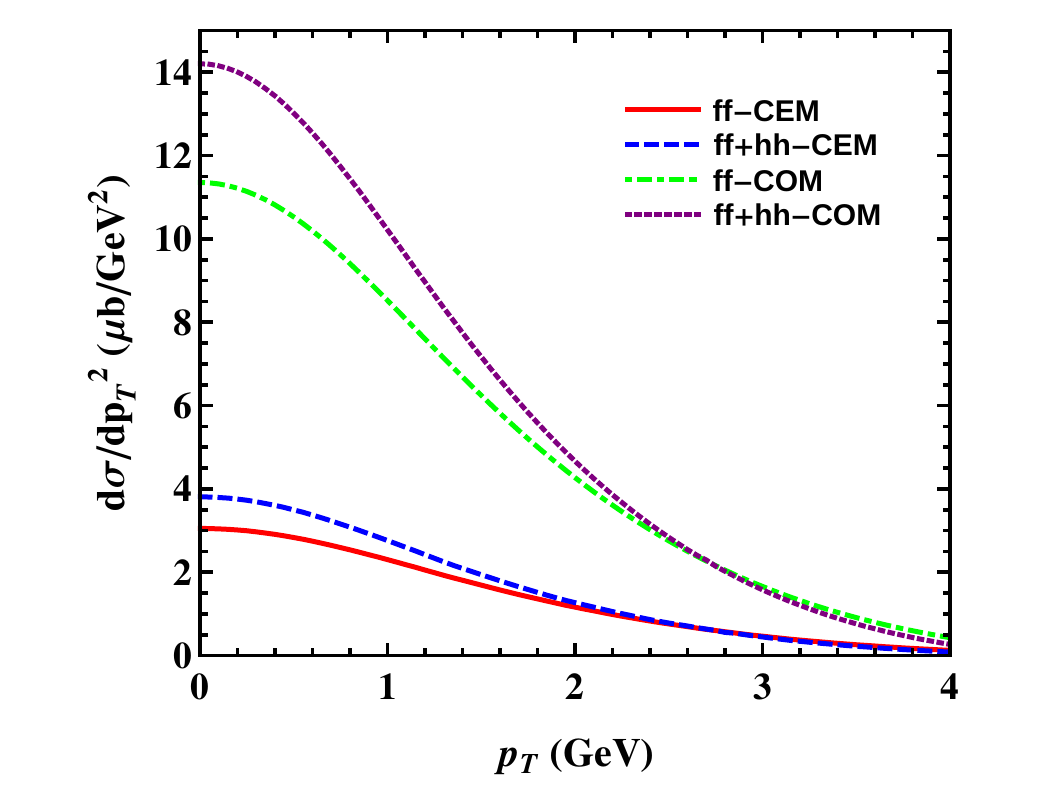}
\hspace{0.1cm}
\small{(b)}\includegraphics[width=7.5cm,height=6cm,clip]{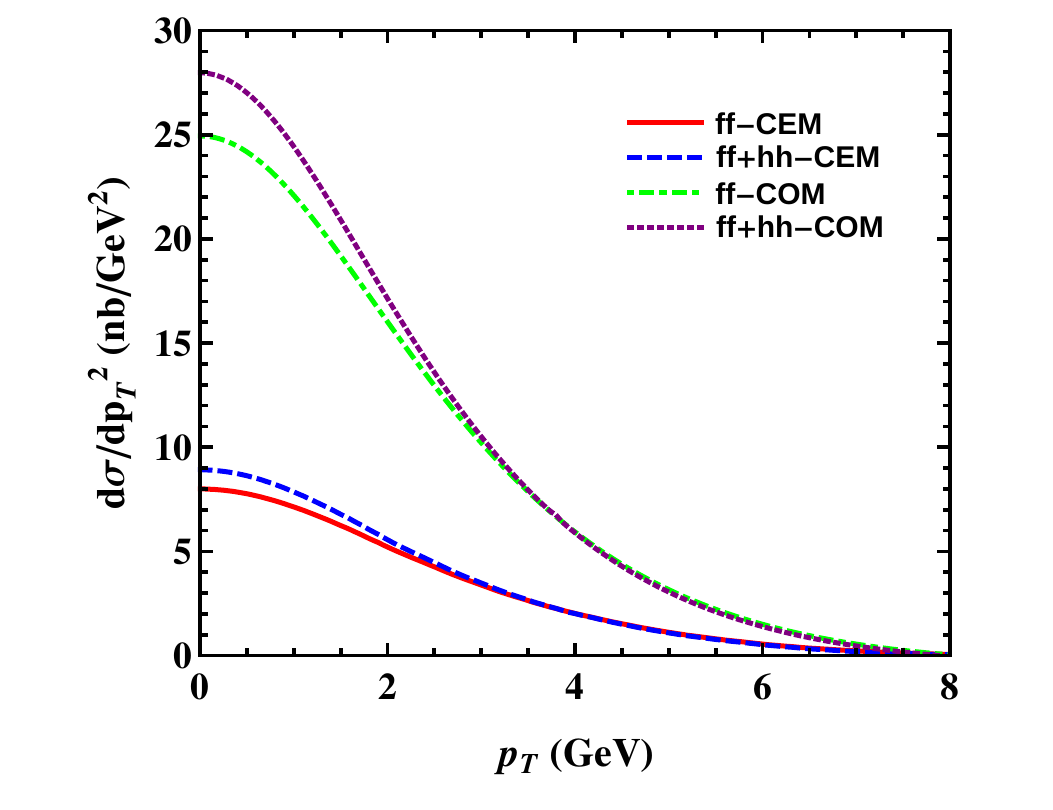}
\end{minipage}
\caption{\label{fig6}(color online). Differential cross section of  (a) $J/\psi$ (left panel)
and  (b) $\Upsilon(1\text{S})$ (right panel) as function of $p_T$ 
in $pp\rightarrow J/\psi(\Upsilon(1\text{S}))+X$ at LHCb ($\sqrt{s}=7$ TeV) energy 
using TMD evolution approach in CEM and COM.  \textquotedblleft Set-I\textquotedblright~LDMEs 
and \textquotedblleft AR\textquotedblright~$R_{NP}$ are used in COM.  The integration range of y is 
$2.0<y<4.5$. The convention in the figure for line styles is same as Fig. \ref{fig3}.}
\end{figure}
\begin{figure}[H]
\begin{minipage}[c]{0.99\textwidth}
\small{(a)}\includegraphics[width=7.5cm,height=6cm,clip]{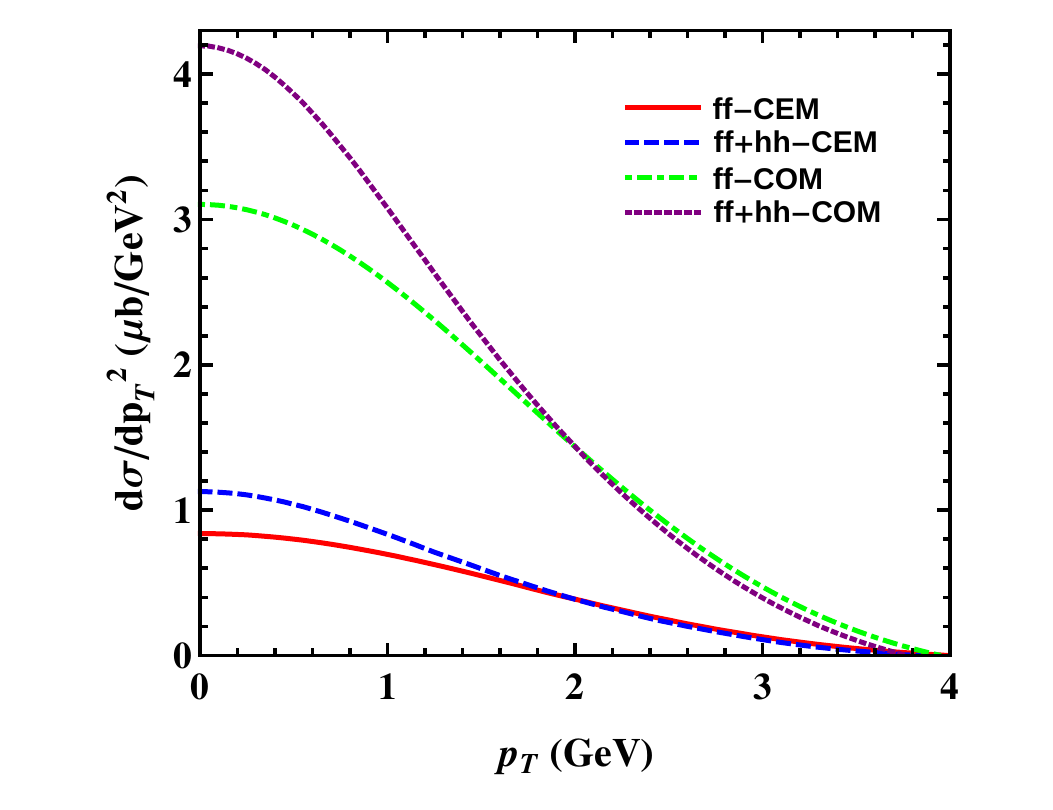}
\hspace{0.1cm}
\small{(b)}\includegraphics[width=7.5cm,height=6cm,clip]{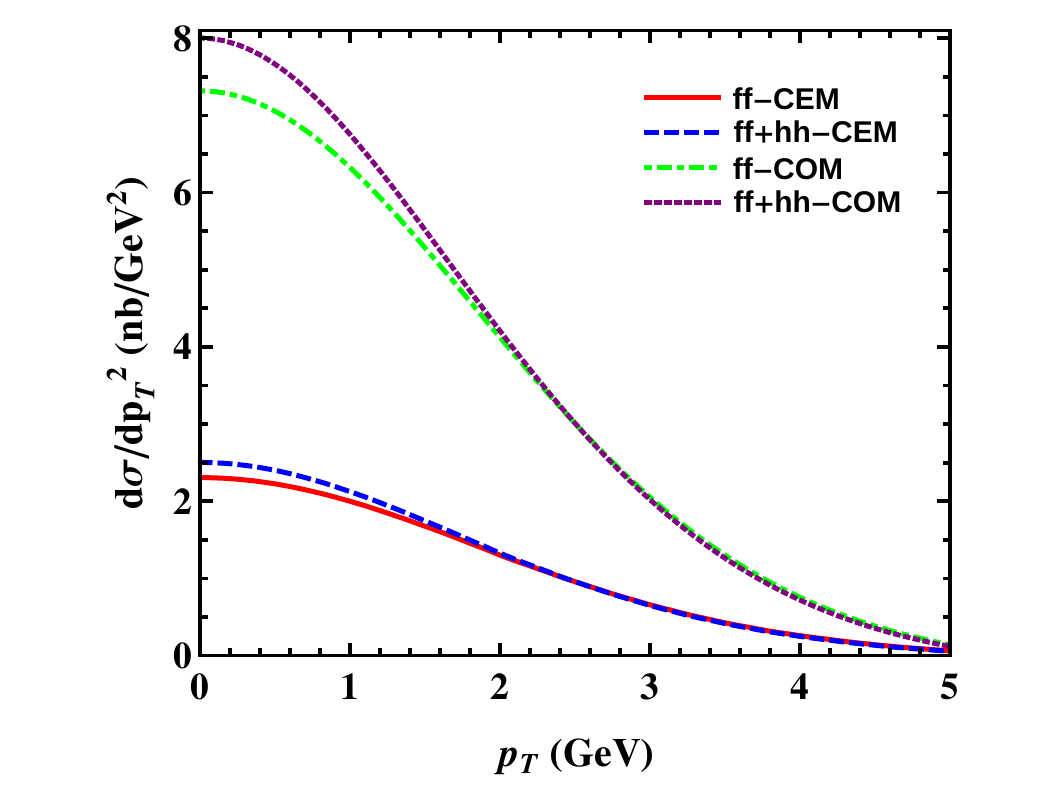}
\end{minipage}
\caption{\label{fig7}(color online). Differential cross section of  (a) $J/\psi$ (left panel)
and  (b) $\Upsilon(1\text{S})$ (right panel) as function of $p_T$ 
in $pp\rightarrow J/\psi(\Upsilon(1\text{S}))+X$ at RHIC ($\sqrt{s}=500$ GeV) energy 
using TMD evolution approach in CEM and COM.  \textquotedblleft Set-I\textquotedblright~LDMEs 
and \textquotedblleft AR\textquotedblright~$R_{NP}$ are used in COM. The integration range of y is 
$-3.0<y<3.0$.  The convention in the figure for
 line styles is same as Fig. \ref{fig3}.}
\end{figure}
\begin{figure}[H]
\begin{minipage}[c]{0.99\textwidth}
\small{(a)}\includegraphics[width=7.5cm,height=6cm,clip]{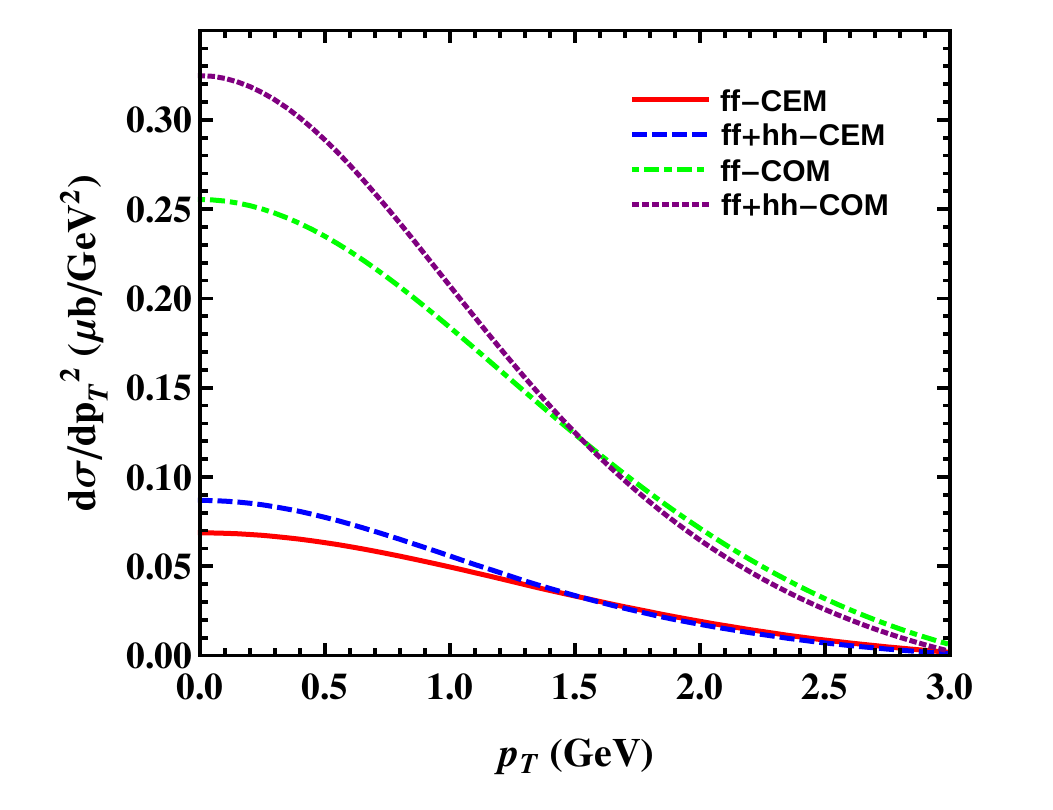}
\hspace{0.1cm}
\small{(b)}\includegraphics[width=7.5cm,height=6cm,clip]{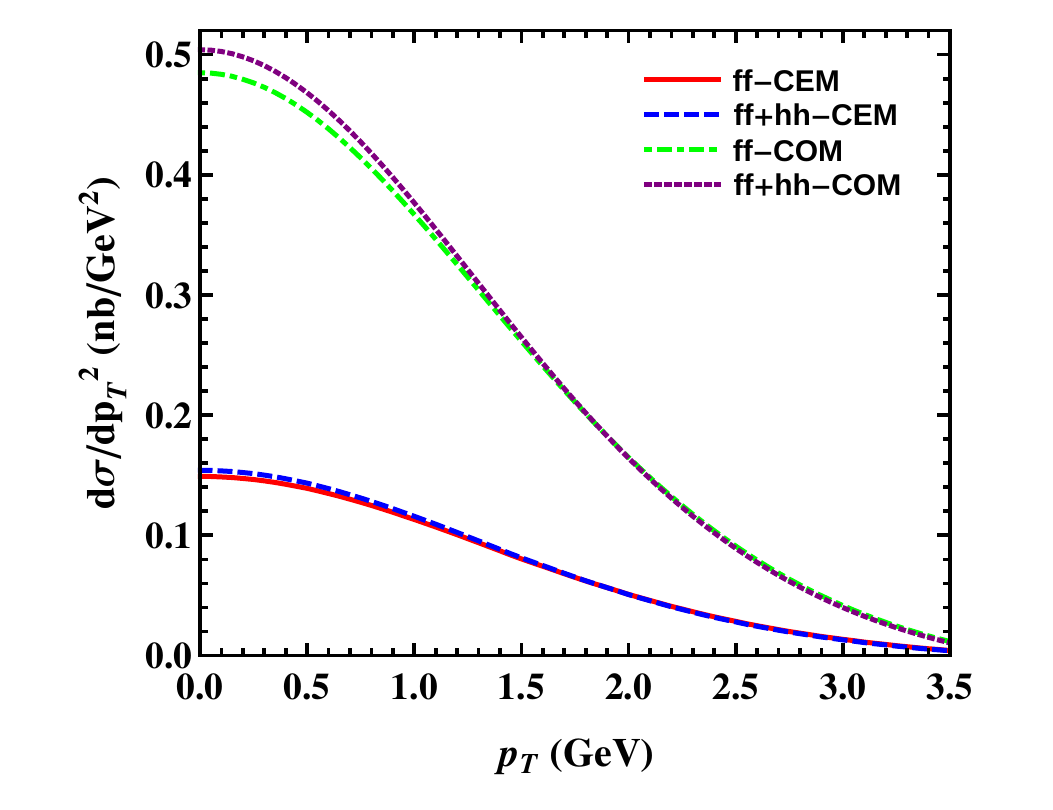}
\end{minipage}
\caption{\label{fig8}(color online). Differential cross section of  (a) $J/\psi$ (left panel)
and  (b) $\Upsilon(1\text{S})$ (right panel) as function of $p_T$ 
in $pp\rightarrow J/\psi(\Upsilon(1\text{S}))+X$ at AFTER ($\sqrt{s}=115$ GeV) energy 
using TMD evolution approach in CEM and COM. \textquotedblleft Set-I\textquotedblright~LDMEs 
and \textquotedblleft AR\textquotedblright~$R_{NP}$ are used in COM. The integration range of y is 
$-0.5<y<0.5$.  The convention in the figure for line styles is same as Fig. \ref{fig3}.}
\end{figure}
\begin{figure}[H]
\begin{minipage}[c]{0.99\textwidth}
\small{(a)}\includegraphics[width=7.5cm,height=6cm,clip]{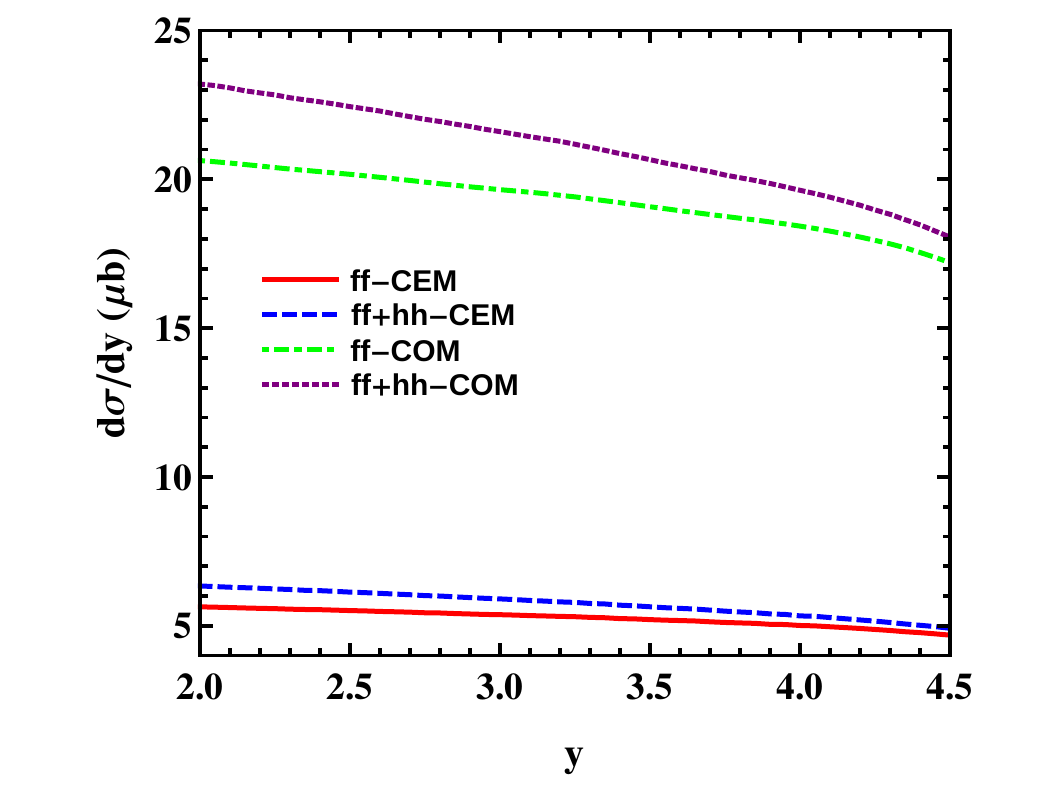}
\hspace{0.1cm}
\small{(b)}\includegraphics[width=7.5cm,height=6cm,clip]{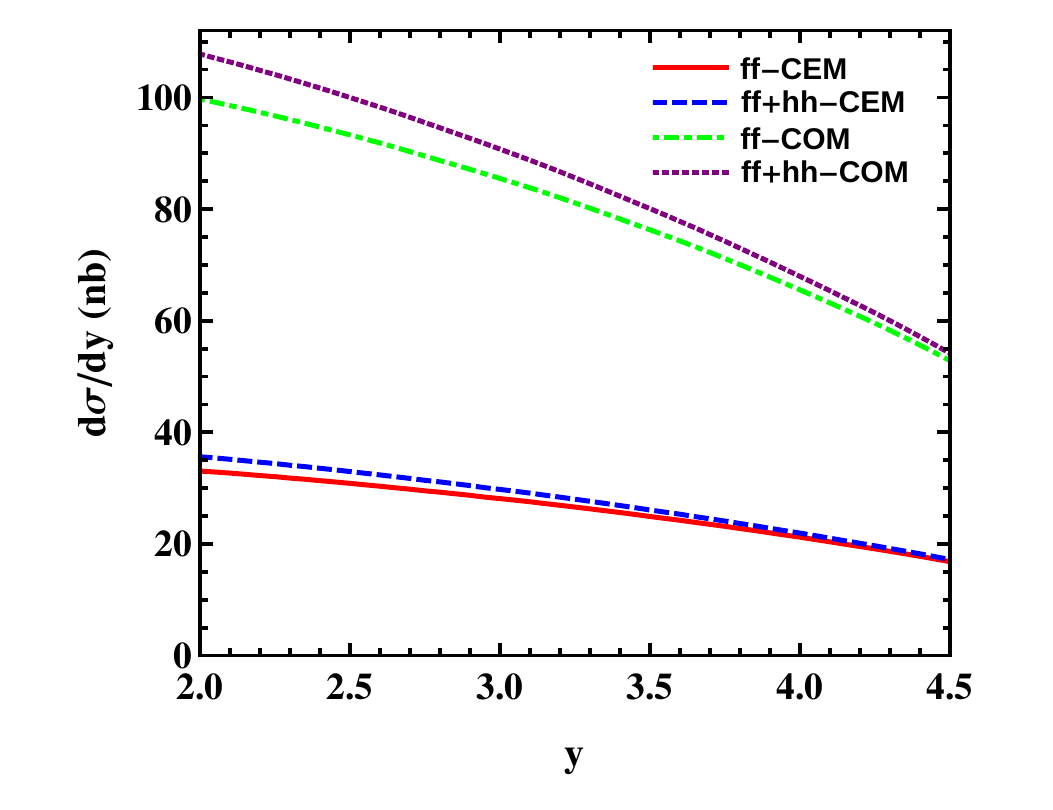}
\end{minipage}
\caption{\label{fig9}(color online). Differential cross section of  (a) $J/\psi$ (left panel)
and  (b) $\Upsilon(1\text{S})$ (right panel) as function of y
in $pp\rightarrow J/\psi(\Upsilon(1\text{S}))+X$ at LHCb ($\sqrt{s}=7$ TeV) energy 
using TMD evolution approach in CEM and COM.  \textquotedblleft Set-I\textquotedblright~LDMEs 
and \textquotedblleft AR\textquotedblright~$R_{NP}$ are used in COM.  The integration range of 
$p_T$ is $0<p_T<4.0$ GeV.  The convention in the figure for line styles is same as Fig. \ref{fig3}.}
\end{figure}
\begin{figure}[H]
\begin{minipage}[c]{0.99\textwidth}
\small{(a)}\includegraphics[width=7.5cm,height=6cm,clip]{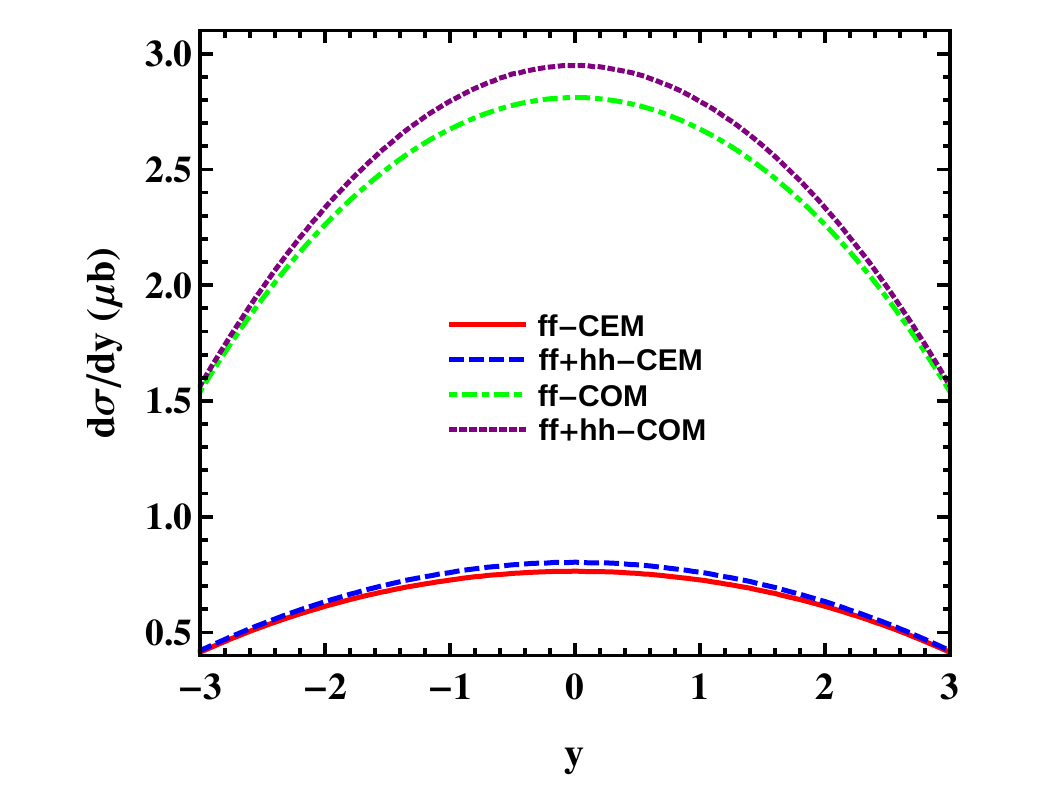}
\hspace{0.1cm}
\small{(b)}\includegraphics[width=7.5cm,height=6cm,clip]{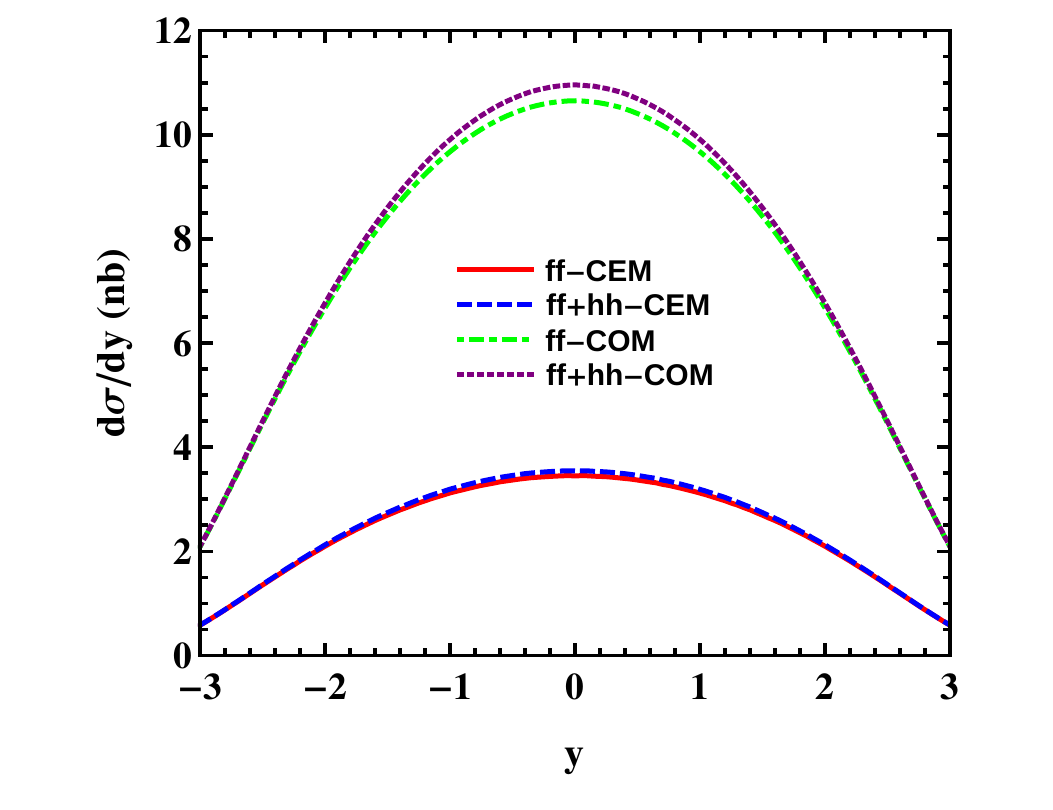}
\end{minipage}
\caption{\label{fig10}(color online). Differential cross section of  (a) $J/\psi$ (left panel)
and  (b) $\Upsilon(1\text{S})$ (right panel) as function of y
in $pp\rightarrow J/\psi(\Upsilon(1\text{S}))+X$ at RHIC ($\sqrt{s}=500$ GeV) energy 
using TMD evolution approach in CEM and COM.  \textquotedblleft Set-I\textquotedblright~LDMEs 
and \textquotedblleft AR\textquotedblright~$R_{NP}$ are used in COM.  The integration range of 
$p_T$ is$0<p_T<4.0$ GeV.  The convention in the figure for line styles is same as Fig. \ref{fig3}.}
\end{figure}
\begin{figure}[H]
\begin{minipage}[c]{0.99\textwidth}
\small{(a)}\includegraphics[width=7.5cm,height=6cm,clip]{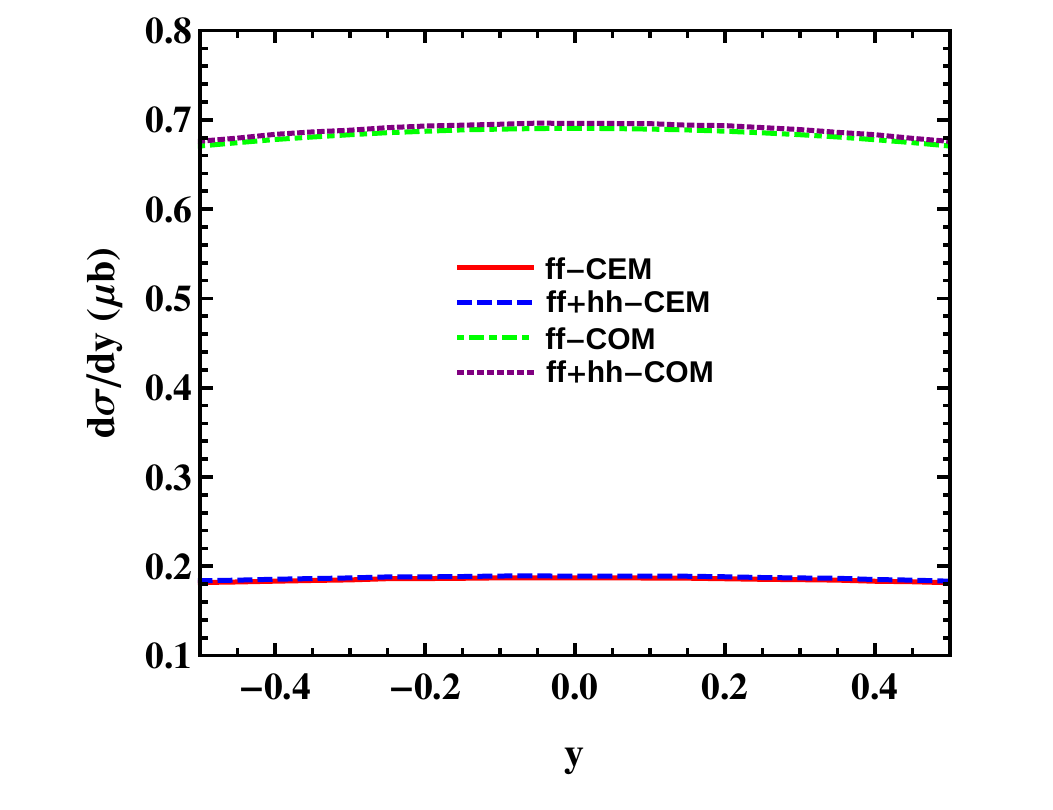}
\hspace{0.1cm}
\small{(b)}\includegraphics[width=7.5cm,height=6cm,clip]{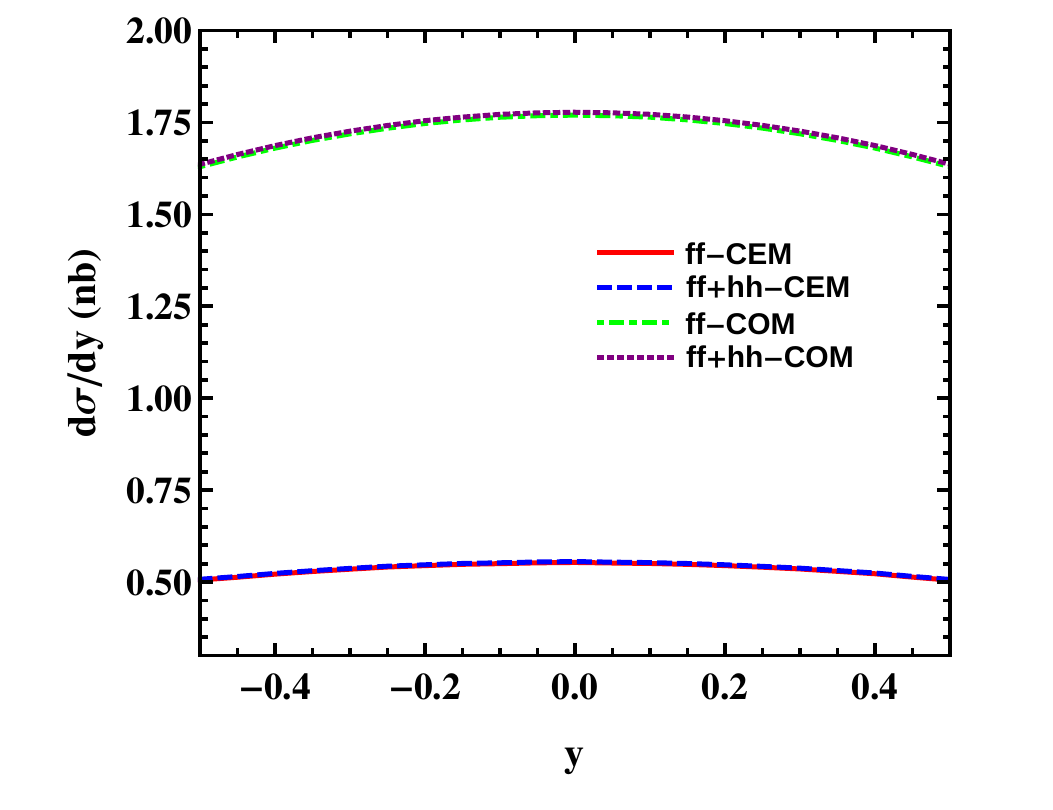}
\end{minipage}
\caption{\label{fig11}(color online). Differential cross section of  (a) $J/\psi$ (left panel)
and  (b) $\Upsilon(1\text{S})$ (right panel) as function of y
in $pp\rightarrow J/\psi(\Upsilon(1\text{S}))+X$ at AFTER ($\sqrt{s}=115$ GeV) energy 
using TMD evolution approach in CEM and COM.  \textquotedblleft Set-I\textquotedblright~LDMEs 
and \textquotedblleft AR\textquotedblright~$R_{NP}$ are used in COM.  The integration range of 
$p_T$ is
$0<p_T<4.0$ GeV.  The convention in the figure for
 line styles is same as \figurename{\ref{fig3}}.}
\end{figure}
\begin{figure}[H]
\begin{minipage}[c]{0.99\textwidth}
\small{(a)}\includegraphics[width=7.5cm,height=6cm,clip]{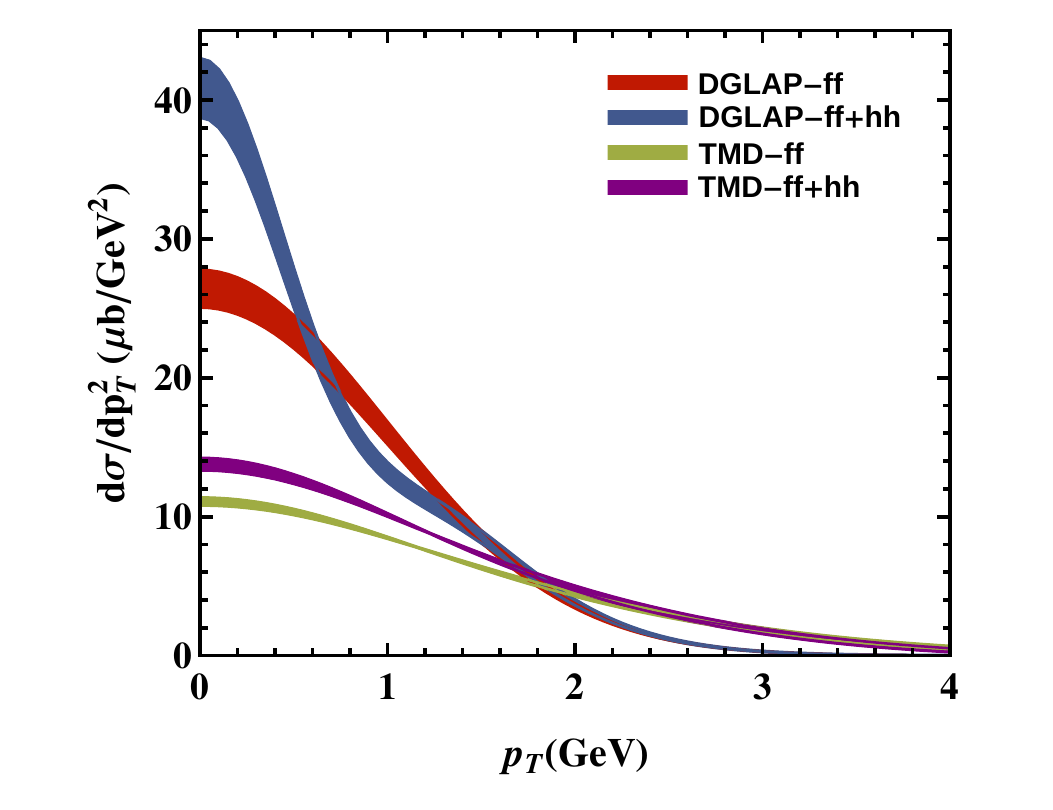}
\hspace{0.1cm}
\small{(b)}\includegraphics[width=7.5cm,height=6cm,clip]{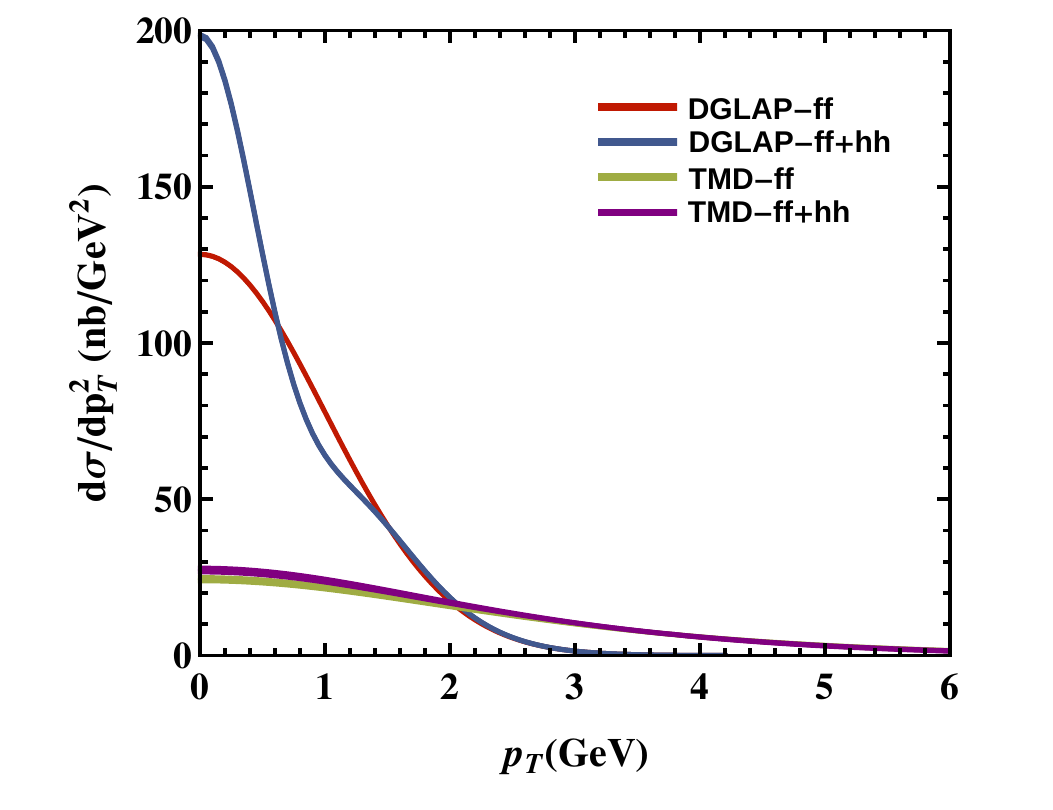}
\end{minipage}
\caption{\label{fig12}(color online).  Differential cross section of (a)  $J/\psi$ (left panel)
and (b)$\Upsilon(1\text{S})$ (right panel) in $pp\rightarrow J/\psi(\Upsilon(1\text{S}))+X$ at LHCb ($\sqrt{s}=7$ TeV)
 in COM using \textquotedblleft Set-I\textquotedblright~LDMEs. For TMD evolution
 \textquotedblleft AR\textquotedblright~$R_{NP}$ is used.  See text for 
the variation of the scale that is shown in
bands. We have chosen $r=\frac{1}{3}$
 and $\langle k^2_{\perp}\rangle=1~\mathrm{GeV}^2$  in Model-I for DGLAP evolution.}
\end{figure}

The transverse momentum distribution is evaluated in model-I from Eq.\eqref{m1ff}  and  Eq.\eqref{m1hh} and
in model-II from Eq.\eqref{m2ff}  and  Eq.\eqref{m2hh}. We have taken two values of the 
Gaussian width $\langle k^2_{\perp}\rangle=0.25$, 1 GeV$^2$ and two values for the parameter
$r=\frac13$, $\frac23$  for the numerical estimation of $p_T$ and $y$ spectra of quarkonium
in DGLAP evolution approach. In \figurename{\ref{fig1}-\ref{fig12}},  \textquotedblleft Set-I\textquotedblright~LDMEs and 
\textquotedblleft AR\textquotedblright~$R_{NP}$ factor are used for color octet states and TMD evolution respectively. 
$p_T$ spectrum of quarkonium is normalized with
total cross section in \figurename{\ref{fig1}} and \figurename{\ref{fig2}} in model-I and model-II resulting in the 
cancellation of scale dependent terms. As a result, 
$p_T$ spectrum of quarkonium is independent of center of mass energy and quarkonium mass as shown in
\figurename{\ref{fig1}} and \figurename{\ref{fig2}}.
The quarkonium $p_T$ spectra presented in \figurename{\ref{fig1}} and \figurename{\ref{fig2}} agree with  that 
we obtained in Ref. \cite{Mukherjee:2015smo} using CEM. The contribution of linearly polarized gluons in $p_T$ 
integrated cross section is zero. Noticeable modifications in the quarkonium $p_T$ distribution are observed upon taking
the linearly polarized gluons into consideration along with the unpolarized gluons, in the 
scattering process. The effect of linearly polarized gluons on  the $p_T$ spectrum of $J/\psi$ and $\Upsilon(1\text{S})$
is limited to low $p_T<0.5$ GeV. Model II gives higher values of the normalized cross section compared to model I. \par
Rapidity distribution of $J/\psi$ and $\Upsilon(1\text{S})$ is shown in \figurename{\ref{fig3}-\ref{fig5}} and is
estimated in model-I for $\langle k^2_{\perp}\rangle=1$ GeV$^2$ and  $r=\frac13$. Rapidity distribution
is obtained by integrating $p_T$ $\in$ [0, 0.5 GeV]. The small window of $p_T$ bin ($0<p_T<0.5$)
is chosen to illustrate the effect of linearly polarized gluons in unpolarized $pp$ collision.
The rapidity spectrum of quarkonium obtained in COM is compared with that of CEM \cite{Mukherjee:2015smo} which is shown 
in the same figures, for model-I.
Comparatively, the production rates of $J/\psi$ and $\Upsilon(1\text{S})$ are slightly higher in COM than CEM. The rapidity 
distribution of quarkonium in model-II also follows the pattern obtained through model-I, however with less magnitude. The $p_T$ and $y$ distribution of quarkonium increase with increasing 
$k_{\rm max} (= \sqrt{\langle k_{\perp a}^2\rangle}$) in model-II. The rapidity distribution is enhanced by 
inclusion of linearly polarized gluons in quarkonium production. 
The enhancement in the rapidity distribution is more at LHCb compared
to RHIC and AFTER experiments.\par
The quarkonium production through COM (\textquotedblleft Set-I\textquotedblright~LDMEs) and CEM within TMD evolution approach
using  \textquotedblleft AR\textquotedblright~$R_{NP}$ factor are compared in \figurename{\ref{fig6}-\ref{fig11}}.
$p_T$ distribution of quarkonium in TMD evolution approach is shown in \figurename{\ref{fig6}-\ref{fig8}} at
LHCb, RHIC and AFTER energies using Eqs.\eqref{tmdevo}-\eqref{evoleq2}. The effect of linearly polarized
gluons increases with center of mass energy of the process. The $p_T$ distribution of $J/\psi$ is  greatly 
affected by linearly polarized gluons at LHCb energy compared to RHIC and AFTER energies. Nevertheless, the 
effect is sizable at low $p_T$. The effect of linearly polarized gluons is less in $\Upsilon(1\text{S})$ production
due to $\Upsilon(1\text{S})$ mass. 
The rapidity spectrum of quarkonium is shown in \figurename{\ref{fig9}-\ref{fig11}} using TMD evolution approach.
 Transverse momentum is integrated in the range of $0<p_T<4$ GeV for $y$ distribution.
In TMD evolution, the production rates of $J/\psi$ and $\Upsilon(1\text{S})$ are more in COM. However, the effect of 
$h_1^{\perp g}$ in $J/\psi$ production is significantly high in COM compared with CEM.\par
\begin{figure}[H]
\begin{minipage}[c]{0.99\textwidth}
\small{(a)}\includegraphics[width=7.5cm,height=6cm,clip]{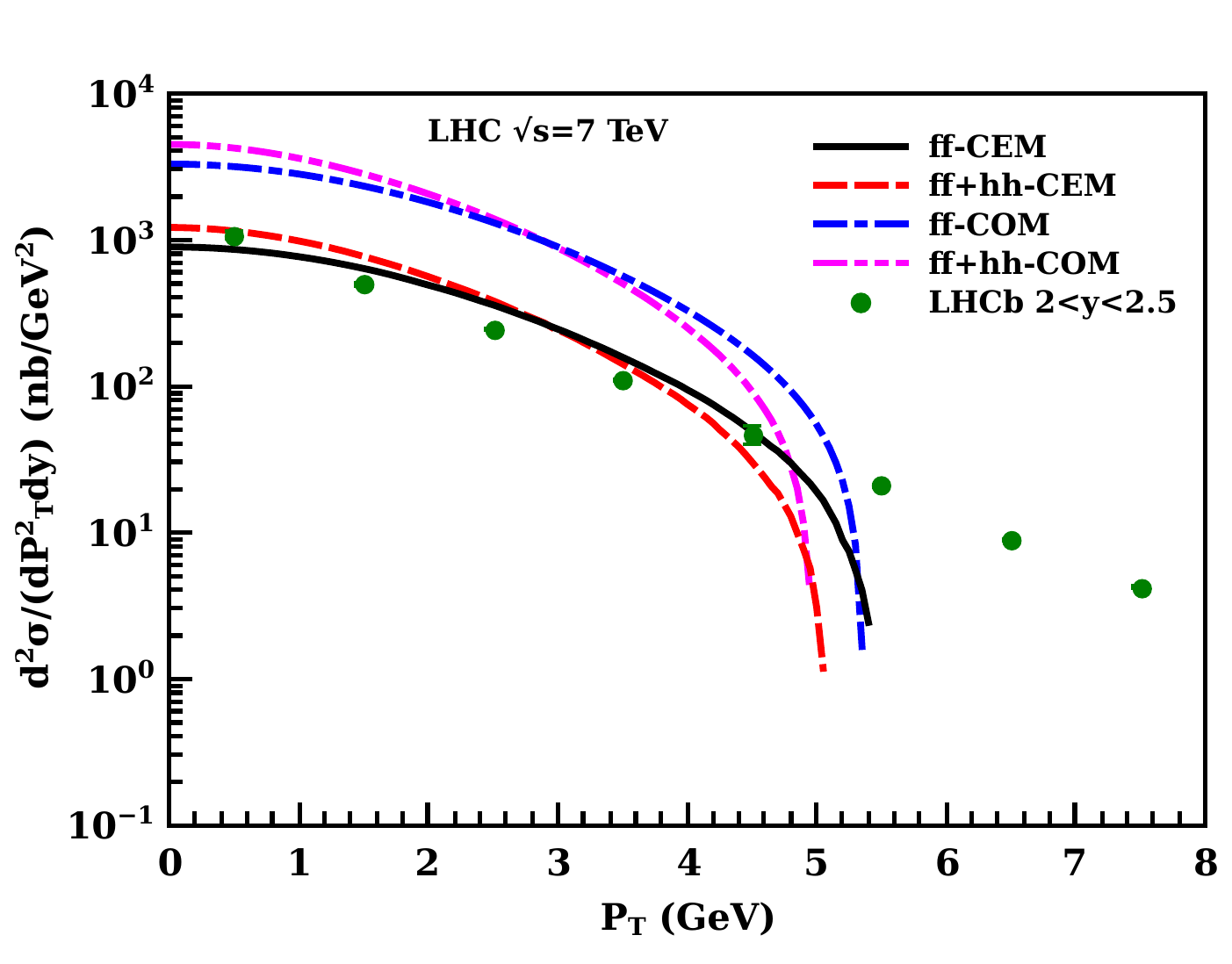}
\hspace{0.1cm}
\small{(b)}\includegraphics[width=7.5cm,height=6cm,clip]{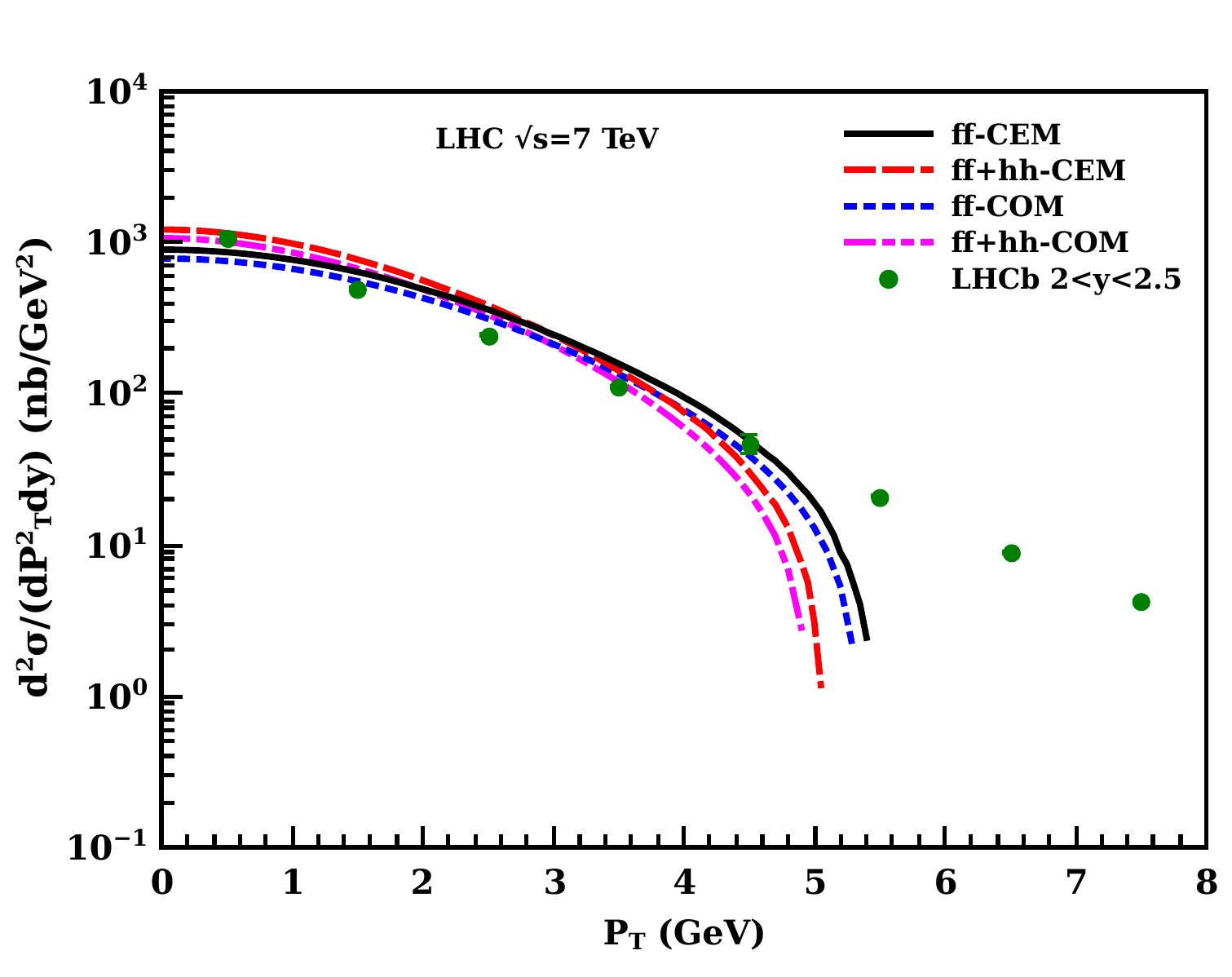}
\end{minipage}
\caption{\label{fig13}(color online). Differential cross section of $J/\psi$  at LHCb ($\sqrt{s}=7$ 
TeV) as function of $p_T$ in $pp\rightarrow J/\psi+X$ using  (a) \textquotedblleft 
Set-I\textquotedblright~(left) and (b)  \textquotedblleft Set-II\textquotedblright~(right) LDMEs in 
COM  within TMD evolution approach for  \textquotedblleft AR\textquotedblright~$R_{NP}$  
factor. Data is taken from \cite{Aaij:2011jh}. The convention in the figure for line styles is same 
as Fig. \ref{fig3}. The rapidity in the range $2.0<y<2.5$   is chosen.}
\end{figure}
\begin{figure}[H]
\begin{minipage}[c]{0.99\textwidth}
\small{(a)}\includegraphics[width=7.5cm,height=6cm,clip]{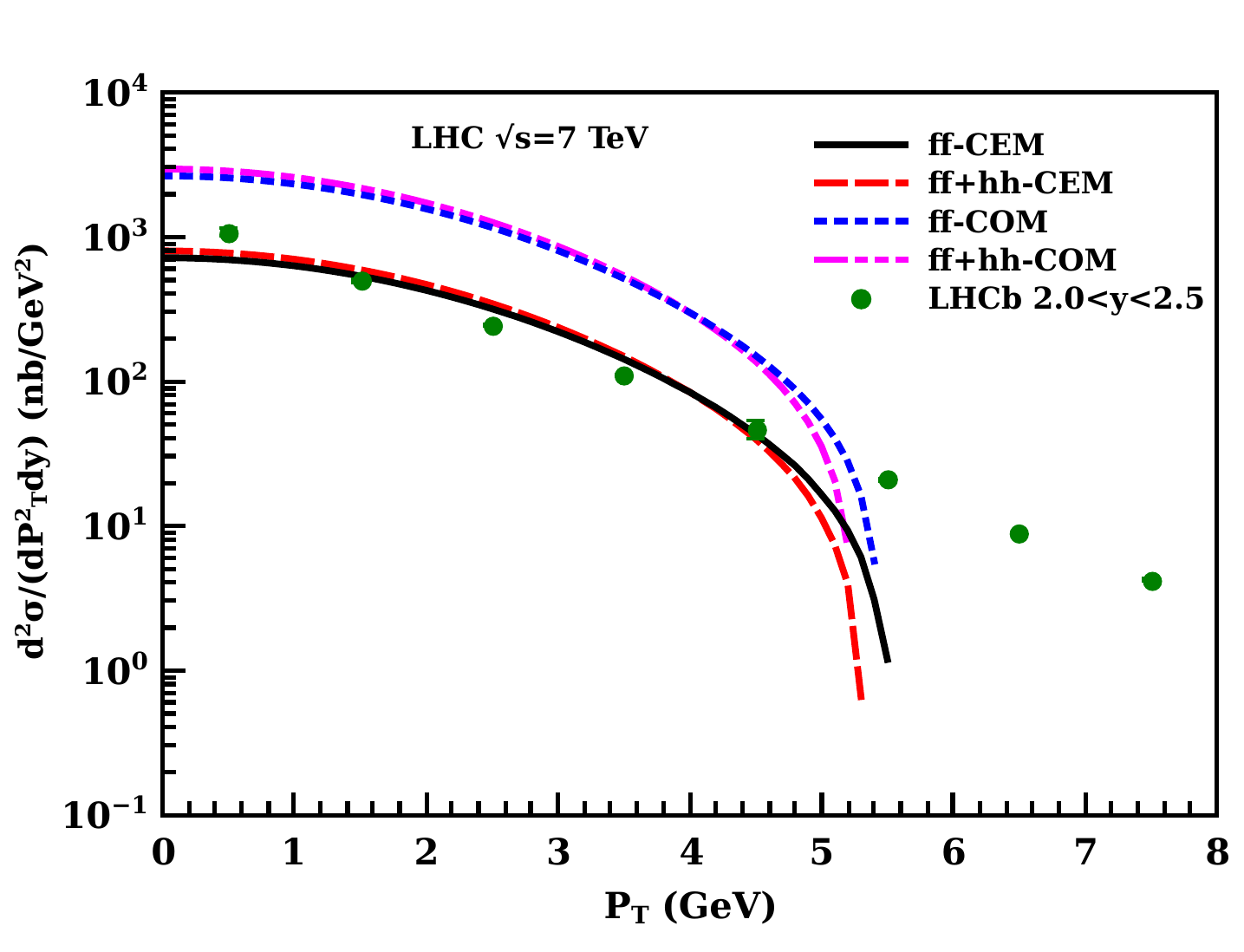}
\hspace{0.1cm}
\small{(b)}\includegraphics[width=7.5cm,height=6cm,clip]{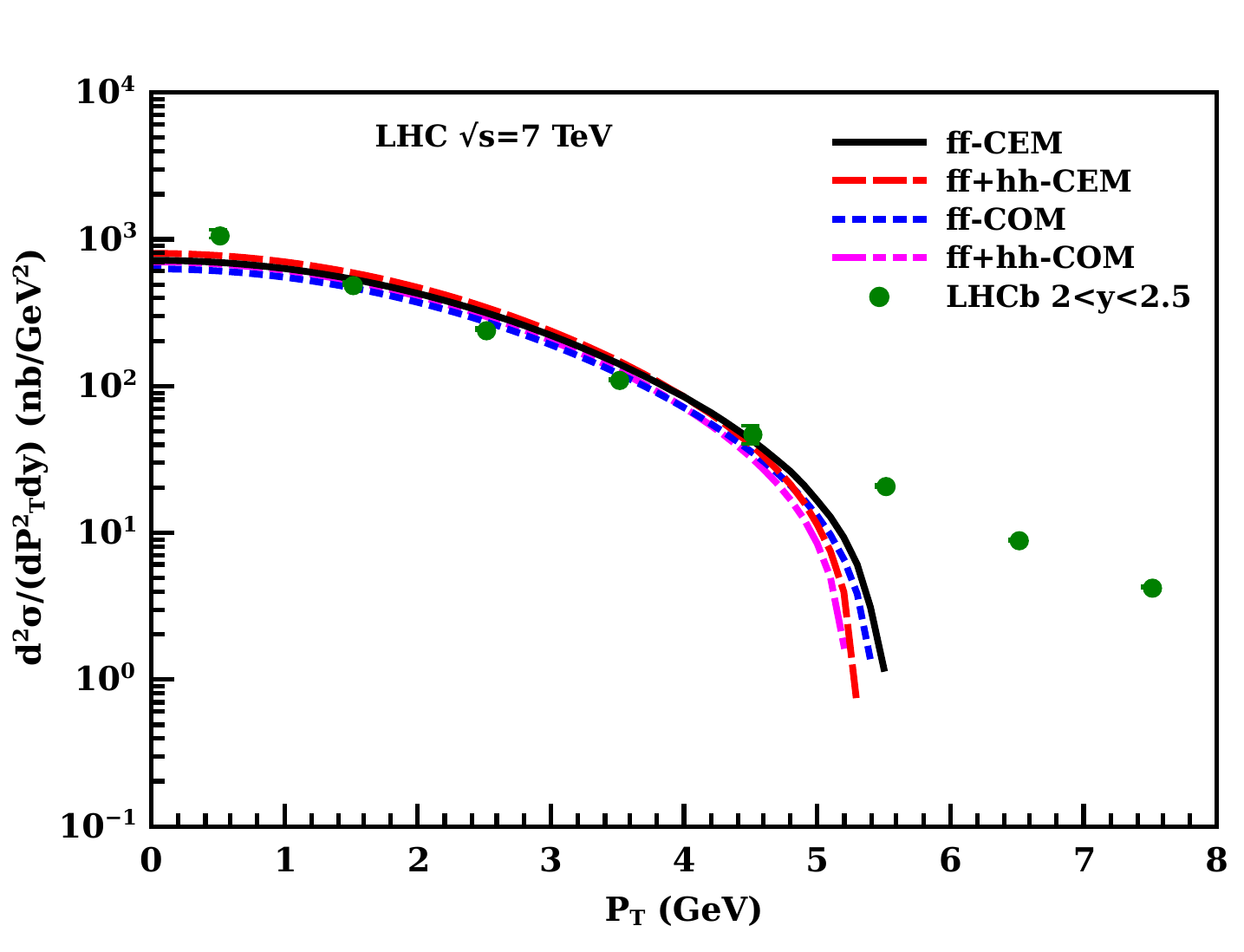}
\end{minipage}
\caption{\label{fig14}(color online). Differential cross section of $J/\psi$  at LHCb ($\sqrt{s}=7$ 
TeV) as function of $p_T$ in $pp\rightarrow J/\psi+X$ using  (a) \textquotedblleft 
Set-I\textquotedblright~(left) and (b)  \textquotedblleft Set-II\textquotedblright~(right) LDMEs in 
COM  within TMD evolution approach for  \textquotedblleft BLNY\textquotedblright~$R_{NP}$  
factor. Data is taken from \cite{Aaij:2011jh}. The convention in the figure for line styles is same 
as Fig. 
\ref{fig3}. The rapidity in the range $2.0<y<2.5$   is chosen.}
\end{figure}
\begin{figure}[H]
\begin{minipage}[c]{0.99\textwidth}
\small{(a)}\includegraphics[width=7.5cm,height=6cm,clip]{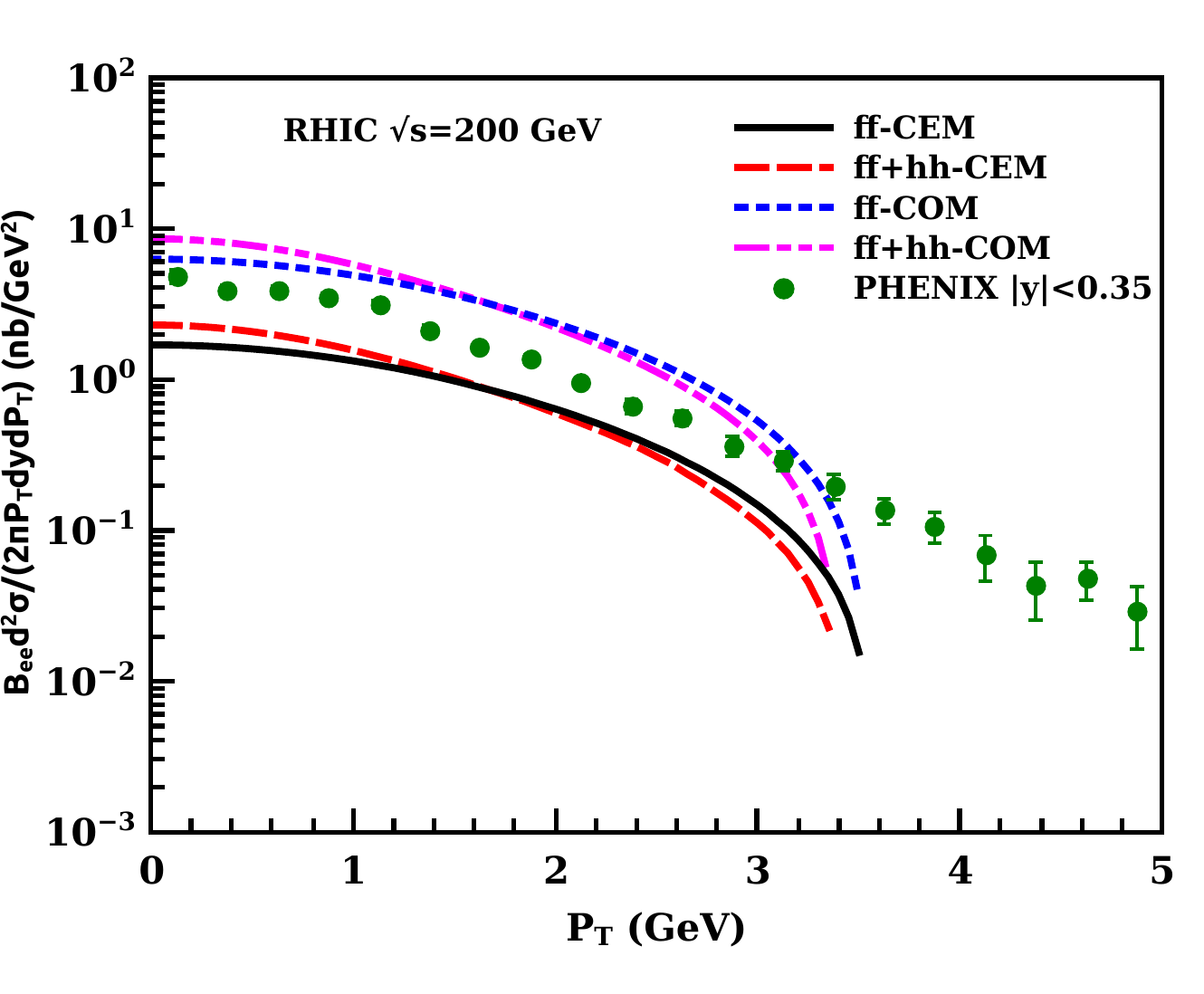}
\hspace{0.1cm}
\small{(b)}\includegraphics[width=7.5cm,height=6cm,clip]{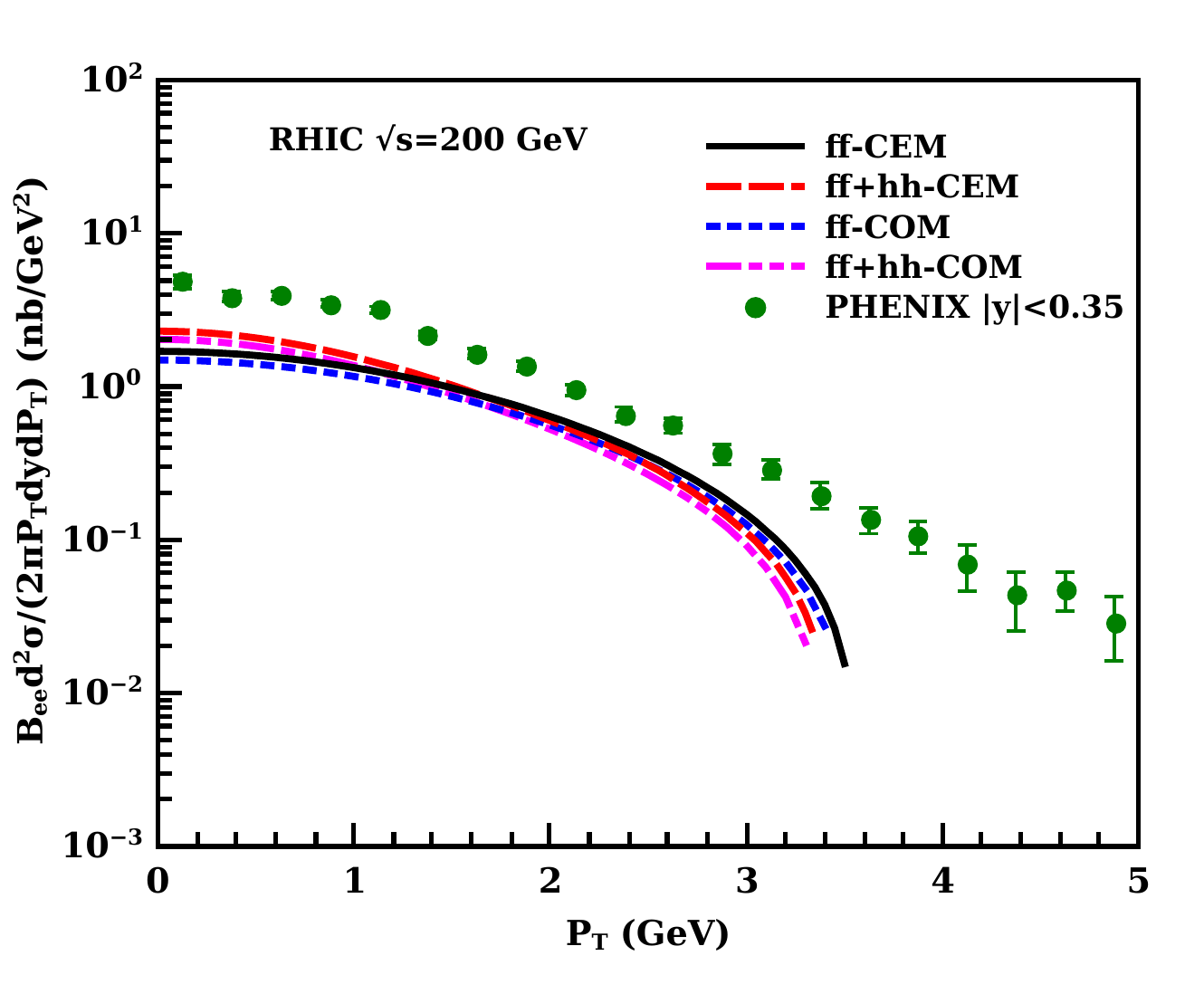}
\end{minipage}
\caption{\label{fig15}(color online). Differential cross section of $J/\psi$  at RHIC 
($\sqrt{s}=200$ GeV) as function of $p_T$ in $pp\rightarrow J/\psi+X$ using  (a) \textquotedblleft 
Set-I\textquotedblright~(left) and (b)  \textquotedblleft Set-II\textquotedblright~(right) LDMEs in 
COM  within TMD evolution approach for  \textquotedblleft AR\textquotedblright~$R_{NP}$  
factor. 
Data is taken from \cite{Adare:2009js}. The convention in the figure for line styles is same as 
Fig. \ref{fig3}. The rapidity in the range $-0.35<y<0.35$   is chosen.}
\end{figure}
\begin{figure}[H]
\begin{minipage}[c]{0.99\textwidth}
\small{(a)}\includegraphics[width=7.5cm,height=6cm,clip]{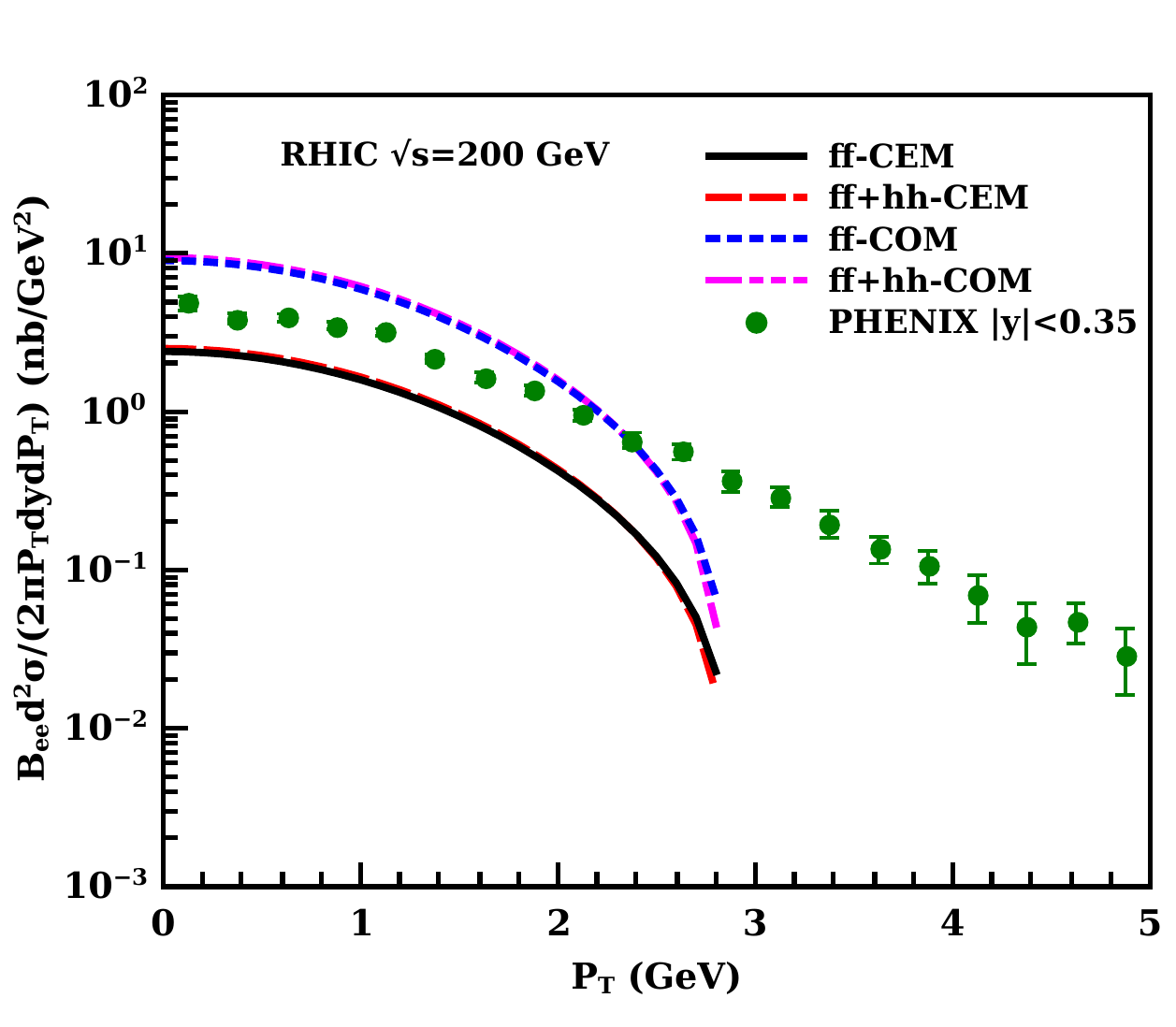}
\hspace{0.1cm}
\small{(b)}\includegraphics[width=7.5cm,height=6cm,clip]{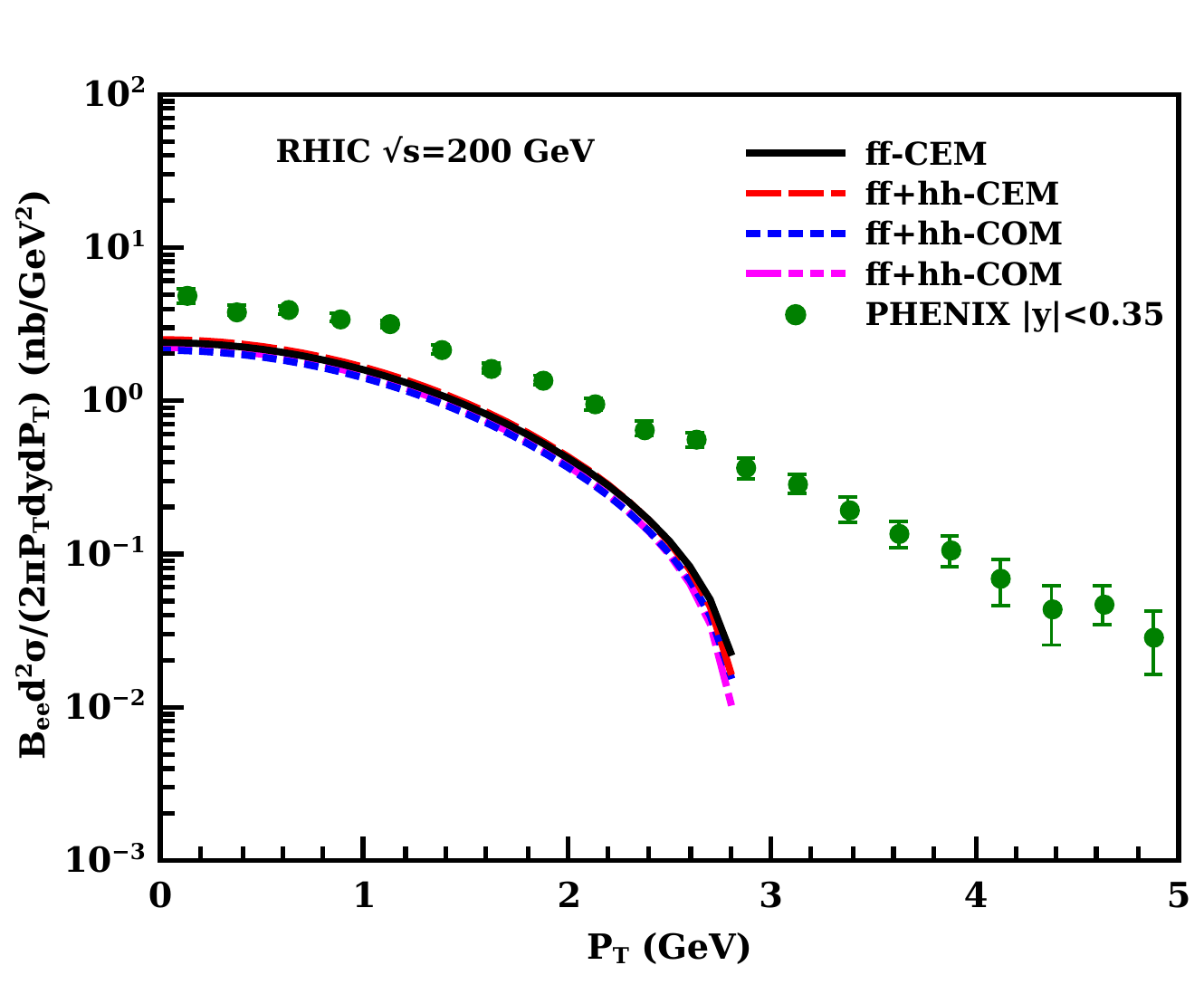}
\end{minipage}
\caption{\label{fig16}(color online). Differential cross section of $J/\psi$  at RHIC 
($\sqrt{s}=200$ GeV) as function of $p_T$ in $pp\rightarrow J/\psi+X$ using  (a) \textquotedblleft 
Set-I\textquotedblright~(left) and (b)  \textquotedblleft Set-II\textquotedblright~(right) LDMEs in 
COM within TMD evolution approach for  \textquotedblleft BLNY\textquotedblright~$R_{NP}$  
factor. 
Data is taken from \cite{Adare:2009js}. The convention in the figure for line styles is same as 
Fig. \ref{fig3}. The rapidity in the range $-0.35<y<0.35$   is chosen.}
\end{figure}
\begin{figure}[H]
\begin{minipage}[c]{0.99\textwidth}
\small{(a)}\includegraphics[width=7.5cm,height=6cm,clip]{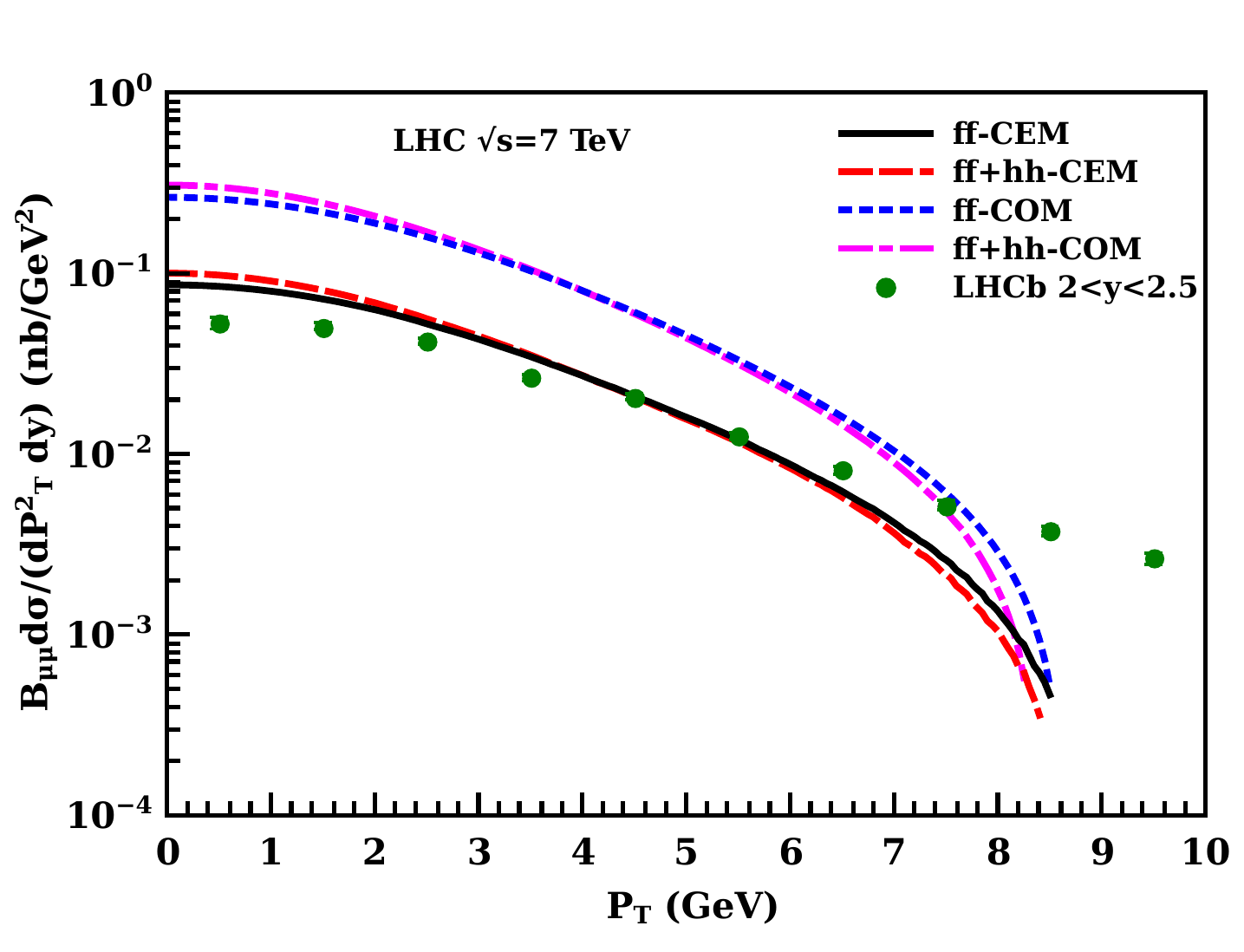}
\hspace{0.1cm}
\small{(b)}\includegraphics[width=7.5cm,height=6cm,clip]{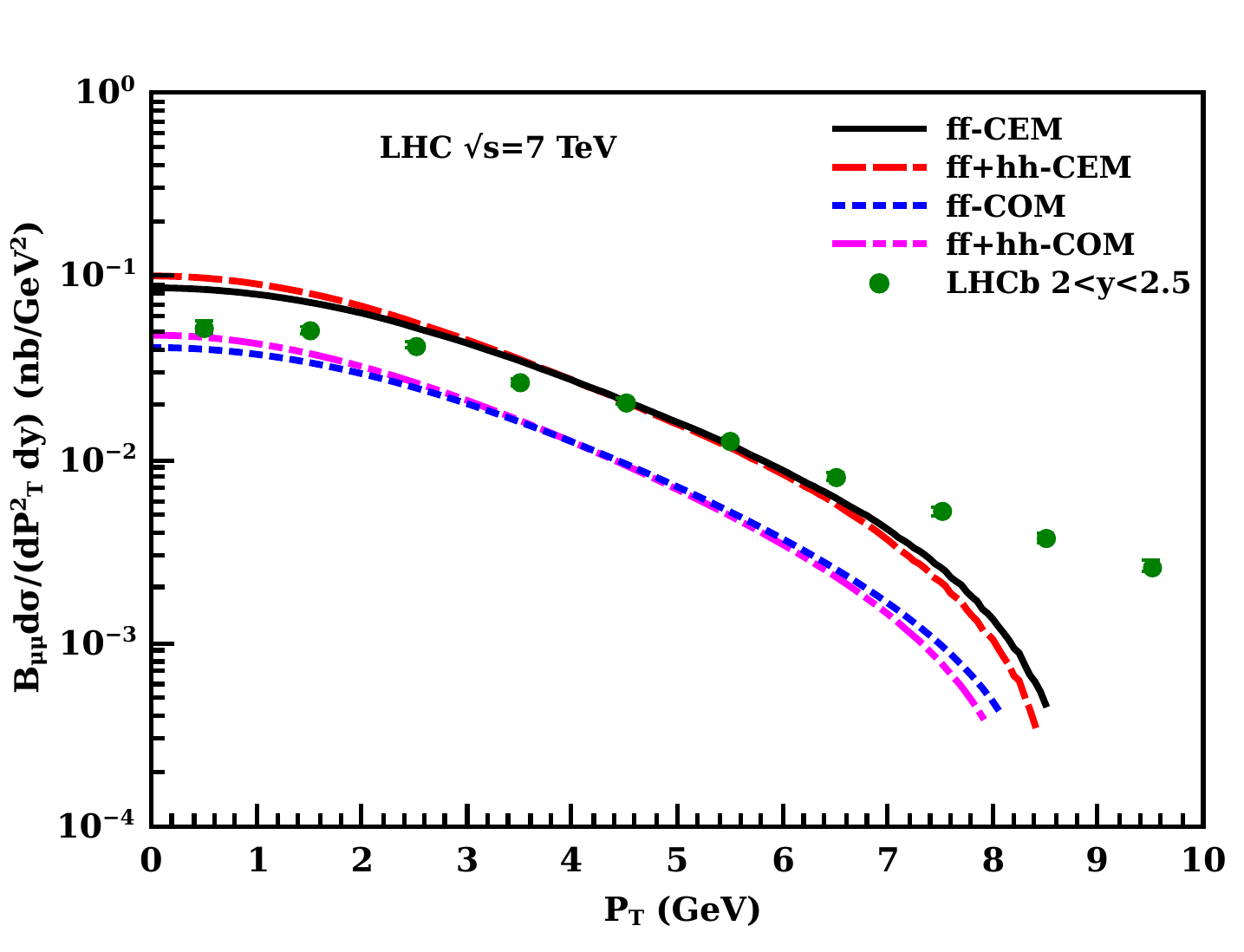}
\end{minipage}
\caption{\label{fig17}(color online). Differential cross section of $\Upsilon(1\text{S}))$  at LHCb 
($\sqrt{s}=7$ TeV) as function of $p_T$ in $pp\rightarrow \Upsilon(1\text{S}))+X$ using  (a) 
\textquotedblleft Set-I\textquotedblright~(left) and (b)  \textquotedblleft 
Set-II\textquotedblright~(right) LDMEs in COM  within TMD evolution approach for  
\textquotedblleft AR\textquotedblright~$R_{NP}$  factor. Data is taken from \cite{LHCb:2012aa}. The 
convention in the figure for line styles is same as Fig. \ref{fig3}. The rapidity in the range 
$2<y<2.5$   is chosen.}
\end{figure}
\begin{figure}[H]
\begin{minipage}[c]{0.99\textwidth}
\small{(a)}\includegraphics[width=7.5cm,height=6cm,clip]{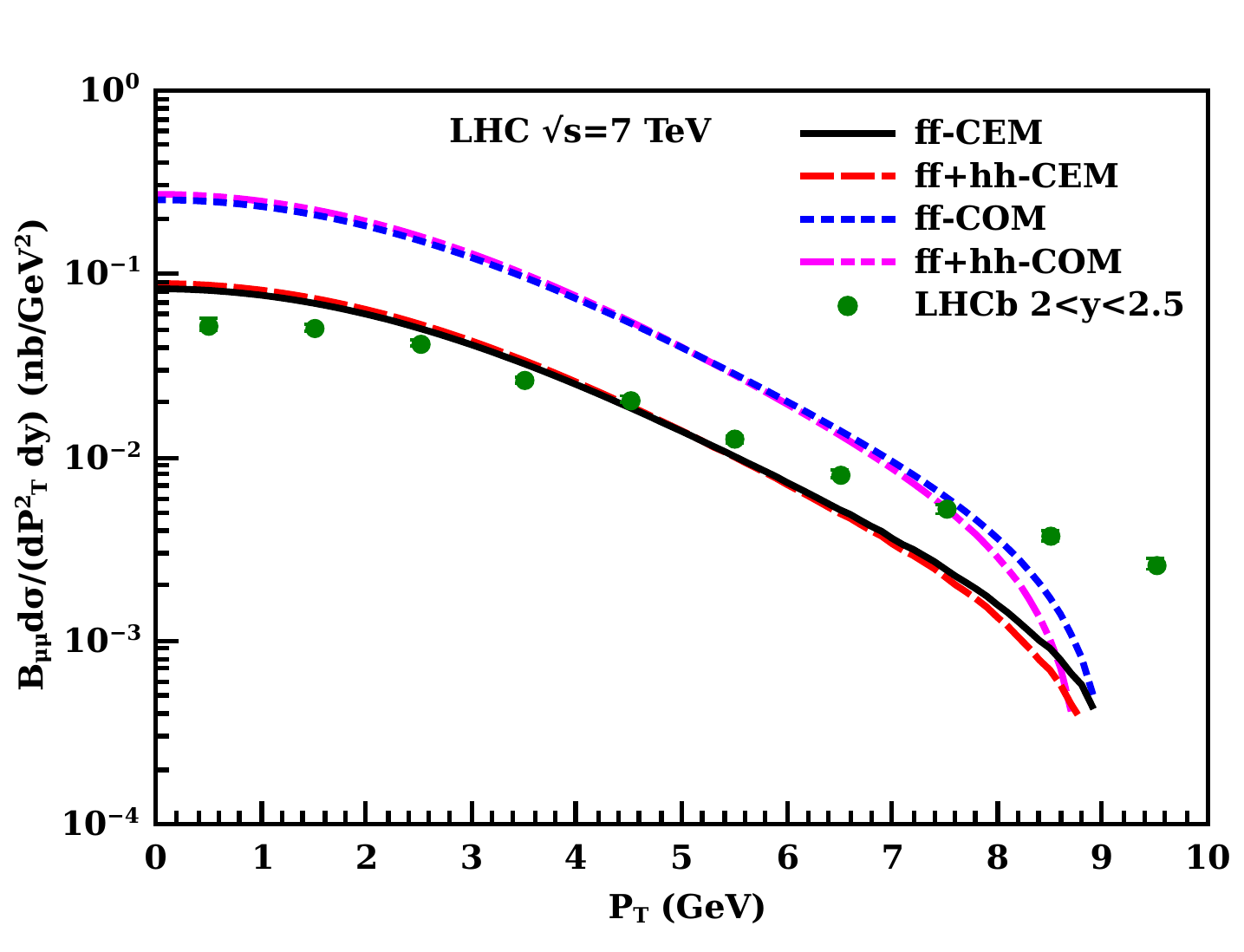}
\hspace{0.1cm}
\small{(b)}\includegraphics[width=7.5cm,height=6cm,clip]{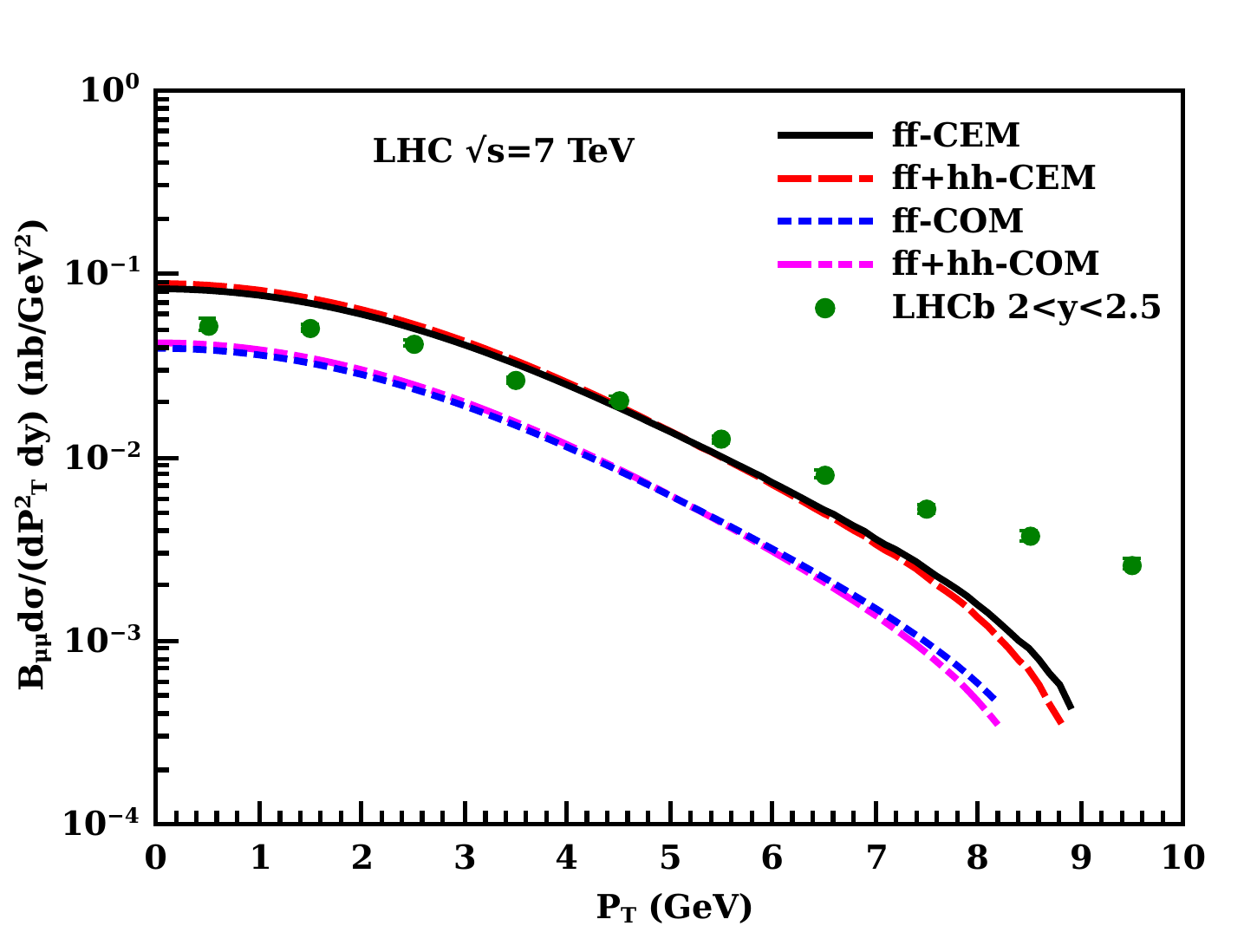}
\end{minipage}
\caption{\label{fig18}(color online). Differential cross section of $\Upsilon(1\text{S}))$  at LHCb 
($\sqrt{s}=7$ TeV) as function of $p_T$ in $pp\rightarrow \Upsilon(1\text{S}))+X$ using  (a) 
\textquotedblleft Set-I\textquotedblright~(left) and (b)  \textquotedblleft 
Set-II\textquotedblright~(right) LDMEs in COM  within TMD evolution approach for  
\textquotedblleft 
BLNY\textquotedblright~$R_{NP}$  factor. Data is taken from \cite{LHCb:2012aa}. The convention in 
the 
figure for line styles is same as Fig. \ref{fig3}. The rapidity in the range $2<y<2.5$   is chosen.}
\end{figure}

The effect of $h_1^{\perp g}$ in TMD evolution is not as much as DGLAP evolution approach.
The comparison  between DGLAP and TMD evolution in COM is shown in \figurename{\ref{fig12}} for $p_T$ spectrum 
of quarkonium. The bands in the figures are obtained by varying the scale (mass of quarkonium) from $Q=3.096$
GeV to 3.596 GeV and 9.398 GeV to 9.898 GeV for $J/\psi$ and $\Upsilon(1\text{S})$ respectively. The gluon momentum
fraction, $x_g$, is proportional to the mass (M) of quarkonium. Hence, the value of $x_g$ is large for massive 
quarkonium and the gluon PDF decreases very rapidly for large values of $x_g$. Therefore, there is not much effect
on the $p_T$ spectrum of $\Upsilon(1\text{S})$ due to the variation in the scale. Moreover, the effect is less in TMD 
evolution compared to DGLAP evolution. The reason is that in DGLAP evolution the collinear PDFs are probed at 
the scale $Q$ whereas PDFs are measured at the initial scale $c/b_\ast$ in TMD evolution approach.
In \figurename{\ref{fig13}-\ref{fig18}}, $p_T$ spectrum of 
quarkonium obtained in TMD evolution  in CEM and COM 
is compared with RHIC data at $\sqrt{s}=200$ GeV \cite{Adare:2009js} and LHCb data at
$\sqrt{s}=7$ TeV \cite{Aaij:2011jh,LHCb:2012aa}.
For obtaining $p_T$ spectrum of quarkonium \textquotedblleft Set-I\textquotedblright~and 
\textquotedblleft Set-II\textquotedblright~LDMEs are considered for color octet states and two nonperturbative Sudakov factors
\textquotedblleft AR\textquotedblright~and \textquotedblleft BLNY\textquotedblright~ are used in TMD evolution
which is shown in \figurename{\ref{fig13}-\ref{fig18}}. 
The theoretical prediction
of $p_T$ spectrum of $J/\psi$ in CEM and COM using \textquotedblleft Set-II\textquotedblright~LDMEs 
is in considerable agreement with  LHCb  data up to low $p_T$  which is shown in \figurename{\ref{fig13}} and \ref{fig14}, 
whereas it is slightly underestimated for RHIC energy as shown in \figurename{\ref{fig15}} and \ref{fig16}.
The $p_T$ spectrum of $\Upsilon(1\text{S})$ is compared with the LHC data  \cite{LHCb:2012aa} and 
is shown in \figurename{\ref{fig17}} and \ref{fig18}.
The obtained $\Upsilon(1\text{S})$ production rate in CEM is in good agreement with LHC data up to 8 GeV. However, 
the low $p_T$ region is slightly overestimated
in COM using \textquotedblleft Set-I\textquotedblright~LDMEs for both $J/\psi$ and $\Upsilon(1\text{S})$ production in 
particular for LHCb experiment.  
The $p_T$ spectrum of $\Upsilon(1\text{S})$ is slightly underestimated in COM using \textquotedblleft Set-II\textquotedblright~LDMEs.
The obtained  $p_T$ spectrum of $J/\psi$ at RHIC energy using \textquotedblleft BLNY\textquotedblright~$R_{NP}$ is falling somewhat faster than 
\textquotedblleft AR\textquotedblright~nonperturbative  parametrization.
The effect of linearly polarized gluons is very less for \textquotedblleft BLNY\textquotedblright~compared to \textquotedblleft AR\textquotedblright~$R_{NP}$.
In \figurename{\ref{fig15}-\ref{fig18}}, $\mathrm{B}_{ee}$ (0.0594) and $\mathrm{B}_{\mu\mu}$ (0.0248) are
the branching ratios of $J/\psi\to e^+e^-$  and $\Upsilon(1\text{S})\to\mu^+\mu^-$  channels respectively.
$J/\psi$ can also be produced in addition to
the direct production in $pp$ collision, for instance, decay from higher mass excited states ($\psi(2\text{S})$ and $\chi_c$) and decay 
of B-meson. The decay of  $\Upsilon(2\text{S})$ ,$\Upsilon(3\text{S})$ and $\chi_b$ contribute to the 
$\Upsilon(1\text{S})$ production. However, in this paper we have considered only the direct production and the inclusion of 
these feed down contribution to the quarkonium production is beyond the
scope of this paper. Of course, leading order calculation for
quarkonium production is insufficient to explain high $p_T$ data. It would be interesting
to investigate the high 
$p_T$ spectrum of quarkonium at LO plus NLO calculation in TMD formalism.
 Inclusion of the so-called Y-term
\cite{Collins:2016hqq} is also
expected to improve the behavior at high $p_T$.  


\section{Conclusion}\label{sec5}

In summary, we studied the transverse momentum and rapidity distribution of $J/\psi$ and $\Upsilon(1\text{S})$ in 
unpolarized proton-proton collision within non-relativistic QCD based color octet model using TMD 
factorization formalism. The LO color octet states 
 $\leftidx{^{1}}{S}{_0}$, $\leftidx{^{3}}{P}{_0}$ and $\leftidx{^{3}}{P}{_2}$ of the initial heavy quark pair in 
 gluon-gluon fusion channel have been considered for  quarkonium production.
The quarkonium production rates are estimated at LHCb, RHIC and AFTER
 center of mass energies. Significant modulations in transverse momentum spectrum of 
$J/\psi$ are shown, in the low  $p_T$ region, when contribution from linearly polarized gluons 
inside an unpolarized proton is included. The rapidity distribution is enhanced with the inclusion of  $h_1^{\perp g}$.
The obtained production rates of quarkonium in COM are compared with CEM.
 The effect of $h_1^{\perp g}$ in $\Upsilon(1\text{S})$ production is not as  dominant as in $J/\psi$ production. 
  Thus quarkonium production is a very useful process
  to extract both unpolarized and linearly polarized gluon density functions. However, one has to understand the quarkonium production mechanism for a reliable extraction of the TMDs. The theoretical predictions of $J/\psi$ and 
$\Upsilon$ in NRQCD based color octet model 
  are in good agreement with the LHC and RHIC data up to low $p_T$. 

 
\bibliographystyle{apsrev} 
\bibliography{reference}

\end{document}